\documentclass{aa}
\usepackage{graphicx}
\usepackage{txfonts}

\begin{document}
   \title{Hydrodynamic simulations of irradiated secondaries in dwarf novae}

   \titlerunning{Irradiated secondaries in dwarf novae}

   \author{M. Viallet
          \and
          J.-M. Hameury}
   \offprints{}

   \institute{Observatoire Astronomique, Universit\'e Louis Pasteur and CNRS,
              11 rue de l'Universit\'e, 67000 Strasbourg, France\\
              \email{viallet@astro.u-strasbg.fr, hameury@astro.u-strasbg.fr}
             }

   \date{Received ; accepted }

  \abstract
   {Secondary stars in dwarf novae are strongly irradiated during outbursts. It has been argued
   that this could result in an enhancement of the mass transfer rate even though the $L_1$ region
   is shadowed from the primary irradiation by the accretion disc. Previous investigations of
   the possibility of a circulation flow transporting heat from hot regions to $L_1$ gave opposite
   answers.}
   {We investigate numerically the surface flow of irradiated secondaries. We consider the full
   time dependent problem and we account for the two-dimensional nature of the flow.}
   {We use a simple model for the irradiation and the geometry of the secondary star: the
   irradiation temperature is treated as a free parameter and the secondary is replaced by
   a spherical star with a space-dependent Coriolis force that mimics the effect of the Roche
   geometry. The Euler equations are solved in spherical coordinates with the TVD-MacCormack
   scheme.}
   {We show that the Coriolis force leads to the formation of a circulation flow from high
   latitude region to the close vicinity of the $L_1$ point. However no heat can be efficiently
   transported to the $L_1$ region due to the rapid radiative cooling of the hot material as it
   enters the equatorial belt shadowed from irradiation. Under the assumption of hydrostatic
   equilibrium, the Coriolis force could lead to a moderate increase of the mass transfer by
   pushing the gas in the vertical direction in the vicinity of $L_1$, but only during the initial
   phases of the outburst (about 15 -- 20 orbital periods). It remains however possible that
   this assumption breaks up due to the strong surface velocity of the flow transiting
   by $L_1$, of the order of the sound speed. In this case however, a three-dimensional
   approach would then be needed to determine the mass flux leaving the secondary.}
   {We therefore conclude that the Coriolis force does not prevent a flow from the heated
   regions of the secondary towards the $L_1$ region, at least during the initial phases
   of an outburst, but the resulting increase of the mass transfer rate is moderate, and
   it is unlikely to be able to account for the duration of long outbursts.}


   \maketitle
%________________________________________________________________

\section{Introduction}

Dwarf novae (DN) are cataclysmic variables that undergo outbursts,
i.e. a sudden increase of their luminosity by a few magnitudes
lasting for a few days (see e.g. Warner 1995). It is now widely
believed that these outbursts are due to a thermal/viscous
instability of the disc triggered when hydrogen becomes partially
ionized. In the standard model of DN (often referred to as the DIM,
see Cannizzo 1993; Lasota 2001 for reviews), the unstable accretion
disc performs a limit cycle between an outburst phase of high
accretion rate onto the compact object and a quiescence phase of low
accretion rate where the disc re-builds. The role of the irradiation
of the secondary star during an outburst is still a matter of
debate.

During an outburst the secondary is likely to be significantly
affected by the strong increase of the accretion luminosity emitted
by the primary, as the irradiation flux may exceed by a large amount
the intrinsic stellar flux of the secondary (see Smak 2004a). It has
been often argued, but not yet demonstrated, that this could result
in a mass transfer enhancement that could, if significant, be an
essential ingredient of the DIM. For example, it has been suggested
that irradiation induced mass transfer enhancement is the origin of
outburst bimodality (Smak 1999) or that it could play an important
role in the outburst/superoutburst phenomenon of the SU Uma stars
(see Hameury 2000; Hameury et al. 2000; Smak 2000).

However, the $L_1$ point is shadowed by the accretion disk (Sarna
1990) and thus is not under the direct influence of irradiation. The
very existence of a circulation flow transporting heated matter from
the irradiated region toward the Lagrange point is not clear and one
does not know if this results in a substantial increase of the mass
transfer rate.

There are some observational clues (see Vogt 1983; Smak 1995; Smak
2004a) that, in some systems, the mass transfer rate is increased
during an outburst or a superoutburst. These observational arguments
have however been questioned by Osaki \& Meyer (2003), who
furthermore argue on theoretical grounds against the possibility of
such a mass transfer enhancement. First, we do not know how the
accretion flux interacts with the secondary envelope and it is not
known if irradiation can raise significantly the effective
temperature on the secondary. Moreover, even in the case where high
temperature gradients and therefore high pressure gradients are
raised, Osaki \& Meyer (2003) argue that the strong Coriolis force
prevents the formation of a circulation flow transporting heat
toward $L_1$.

\begin{figure}[t] %  figure placement: here, top, bottom, or page
   \centering
   \includegraphics[width=7cm,angle=90]{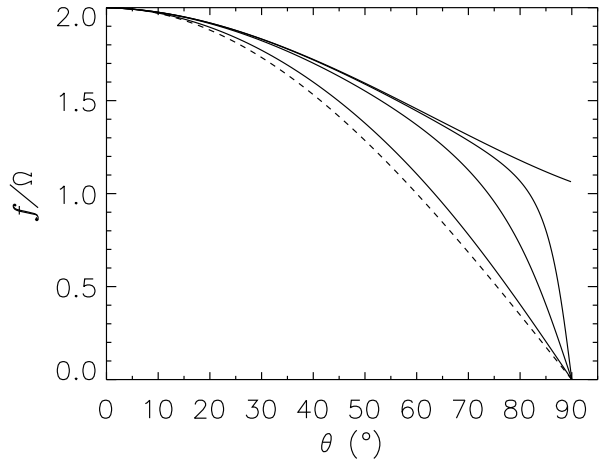}
   \caption{Profile of the Coriolis parameter $f$ along different
   meridians of the Roche lobe: $\phi = 0^\circ$ ($L_1$ meridian),
   $\phi = 5^\circ$, $\phi = 15^\circ$ and $\phi = 45^\circ$ (from top
   to bottom). For comparison the profile corresponding to spherical
   geometry is also shown in dashed line.}
   \label{fmerid}
\end{figure}

Their main argument is that the Coriolis force remains significant
near $L_1$, due to the particular shape of the Roche lobe (contrary
to the spherical case where the Coriolis force vanishes at the
equator); its magnitude remains large enough to deflect any
meridional flow to what is called a \emph{geostrophic flow}, a
steady state where velocity field lines are parallel to isobars (see
Pedlosky 1982). These conclusions were however challenged by
computations by Smak (2004a), showing streamlines converging to the
$L_1$ region. In subsequent papers, Osaki \& Meyer (2004) and Smak
(2004b,c) developed their argumentation without reaching an
agreement.

One must note that the geometric argument from Osaki \& Meyer is a
one-dimensional argument, valid on the $L_1$ meridian only. As the
$L_1$ point is a singular point in the Roche geometry, the strength
of the Coriolis force changes very rapidly as one moves away from
$L_1$ on the equator; this leaves open the possibility that the
heated material does not follow the shortest way along the meridian,
but that it follows an ``easier" way to reach the $L_1$ region. Smak
(2004a) computations were done in a two-dimensional steady state and
thus account for this effect.

The main flaw of both approaches is that the time scale to reach a
steady state could be long, possibly longer than the duration of an
outburst, or that such a steady state could never be reached.

We investigate here the dynamics of the irradiated envelope by using
two-dimensional hydrodynamic simulations. We adopt a very simple
model for the irradiation and geometry of the secondary, but,
contrary to Smak (2004a), we solve the full time-dependent problem.
Our numerical simulations show that the flow can indeed reach the
vicinity of the $L_1$ point, thanks to (and not despite) the
Coriolis force. However, as hot material crosses the boundary
between regions directly illuminated by the primary star and regions
shadowed by the accretion disc, it cools down so rapidly that any
significant heating of the $L_1$ point can not be obtained. If
vertical hydrostatic prevails, no mass transfer enhancement can
occur. However, velocity magnitudes of the order of the sound speed
are obtained in the vicinity of the $L_1$ point and strong vertical
velocity gradients are thus expected, which could lead to turbulence
and/or the breakdown of hydrostatic equilibrium assumption. In this
case, one would need a full 3D analysis of the problem, far beyond
the scope of this paper.

The paper is organized as follow. We present the model in Sect. 
\ref{model} and the numerical code in Sect. \ref{code}. Our
results are given in Sect. \ref{results} and we summarize our
conclusions in Sect. \ref{conclusion}.

\begin{figure}[t] %  figure placement: here, top, bottom, or page
   \centering
   \includegraphics[width=7cm,angle=90]{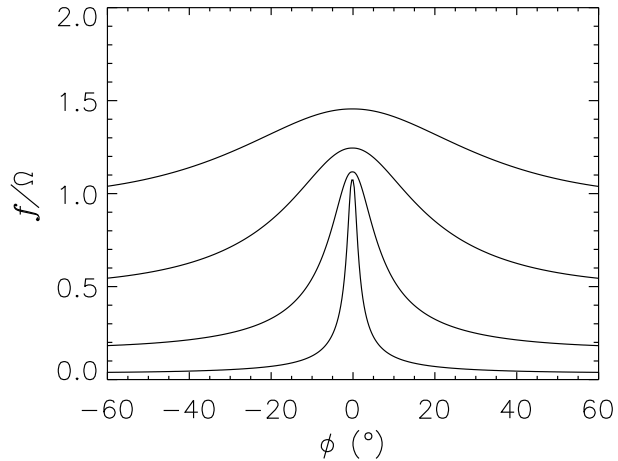}
   \caption{Profile of the Coriolis parameter $f$ on the Roche lobe at
   colatitudes $\theta = 89^\circ$, $\theta = 85^\circ$, $\theta = 75^\circ$
   and $\theta = 60^\circ$ (from bottom to top).}
   \label{flat}
\end{figure}

\section{The model}
\label{model}

We consider three models with characteristics of the three DN
subtypes. The main physical parameters used are given in Table
\ref{params_phys}.

\begin{table*}
\caption{Physical parameters for the three cases investigated in
this paper. $P_\mathrm{orb}$, $a$, $q$ have their usual meanings.
$R_2$ is the mean-radius of the secondary. The secondary surface
temperature $T_\star$ is taken from Smak (2004a). $T_{\rm irr,0}$ is
the maximum irradiation temperature reached during an outburst,
$\Sigma_\mathrm{irr}$  is the column density of the irradiated layer
when $T = T_\mathrm{irr}$. The last column is the height of the
shadow boundary
above the equator when measured on the main meridian.}        % title of Table
\label{params_phys}      % is used to refer this table in the text
\centering                          % used for centering table
\begin{tabular}{c c c c c c c c c c}        % centered columns (4 columns)
\hline \hline                % inserts double horizontal lines
System & Model number & $P_\mathrm{orb}$ (h) & $a\ (R_\odot)$  & $R_2/a$& $q=M_2/M_1$
&$T_\star$ (K) & $T_\mathrm{irr}$ (K) & $\Sigma_\mathrm{irr}$ (g.cm$^{-2}$) & shadow height \\
% table heading
\hline                        % inserts single horizontal line
OY Car & 1 & 1.51 & 0.6 & 0.21 & 0.1 &2500 & 10000 & 350 & $10^\circ$\\ % inserting body of the table
U Gem & 2 & 4.24 & 1.48 & 0.29 &0.36 &3500 & 10000 & 200 & $10^\circ$\\
Z Cam & 3 & 6.96 & 2.17 & 0.35 &0.6 & 4200 & 10000 & 250 & $10^\circ$\\
\hline                                   %inserts single line
\end{tabular}
\end{table*}

\subsection{Geometrical considerations on the dynamics}
\label{geom_cons}

A complete investigation of the problem would require 3D radiative
hydrodynamic simulations of the stellar envelope in the full Roche
potential. In order to be tractable, the problem has to be
simplified. As the scale height of the atmosphere is small compared
to the stellar radius, $h/R \ll 1$ (even in the vicinity of the
$L_1$ point), the vertical dynamical time scale is small compared 
to all other time scales and the vertical hydrostatic equilibrium should be reached 
rapidly. A 2D approach should be sufficient to investigate the important features 
of the surface flow (see however the discussion in Sect. \ref{sectmasstrenh} and in the conclusion).

We therefore solve the full set of the Euler equations on the
surface of the secondary star, whose surface is a Roche
equipotential. As these systems have short orbital periods, the
Coriolis force is an essential dynamical ingredient. In 2D the
Coriolis force enters the equations only via its component parallel
to the surface. Its strength is characterized by the \emph{Coriolis
parameter} $f = 2\vec \Omega . \vec n$ where $\vec n$ is the local
normal to the surface. Fig. \ref{fmerid} shows the profile of $f$
along different meridians of the Roche lobe. On the $L_1$ meridian,
it can be seen that the Coriolis parameter remains large as stated
in Osaki \& Meyer (2003, 2004), but the $\theta$ dependance of $f$
becomes more and more similar to the spherical case as one moves
away from $L_1$. Fig. \ref{flat} show the strong dependence of $f$
on longitude, showing that a 1D calculation is insufficient.

\begin{figure}[t] %  figure placement: here, top, bottom, or page
   \centering
   \includegraphics[width=7cm,angle=90]{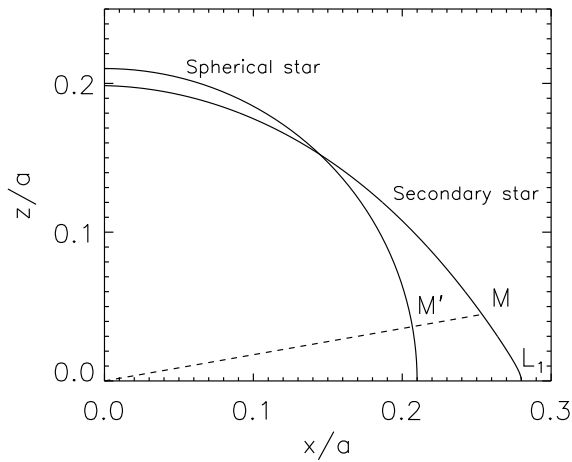}
   \caption{Map $M \rightarrow M'$ from the Roche equipotential to the
   sphere of radius $R_2$ (the mass ratio is $q=0.1$). The figure shows
   a cut along the $L_1$ meridian.}
   \label{mapping}
\end{figure}

The ratio of the inertial force to the Coriolis force  is the
\emph{Rossby number}:

\begin{equation}
\mathfrak{Ro} = \frac{V}{fL},
\end{equation}

\noindent where $V$ and $L$ are characteristic values of the
velocity and length. In our case the relevant scales are $L \sim
R_2$, the mean-radius of the secondary, $V \sim c_\mathrm{s}$, the
sound speed and $f \sim \Omega$, yielding $\mathfrak{Ro} \sim 0.04$
for all models (see Table \ref{params_phys}). As expected, the
Coriolis force dominates the dynamics on the secondary.

One should in principle use Roche coordinates (see Kopal 1969) to
describe the secondary surface. For simplicity, we consider a
spherical star of radius $R_2$ (the mean radius of the Roche lobe)
with a space-dependent Coriolis parameter that mimics the Roche
geometry. This greatly simplifies the problem, as spherical
coordinates are much simpler to use than Roche coordinates. The
mapping between the Roche equipotential and the sphere is
illustrated in Fig. \ref{mapping}; we use a Cartesian system of
coordinates where the center of mass of the secondary is at the
origin, the center of mass of the primary is at $(a,0,0)$ with $a$
the orbital separation and the $z$ axis is perpendicular to the
orbital plane. The spherical approximation is poor in the $L_1$
region, where departure from spherical geometry is the most
important. This has two consequences. First the geometrical terms in
the Euler momentum equation are not well taken into account.
However, in this region, the dynamic is dominated by the Coriolis
force rather than by geometrical effect (note that the Rossby number
is also the ratio of geometric terms to the Coriolis one in the
Euler equations).  Second we underestimate distances: for example
the distance from the shadow boundary to $L_1$ is underestimated by
$\sim 20 - 30 \%$ in our model (see Fig. \ref{mapping}). This is not
a serious difficulty, since it is easy to extrapolate our results to
a real Roche geometry.

The Roche geometry is singular at $L_1$, but we assume that the
stellar surface is at a minimum distance $\Delta r = 0.02 R_2$ below
$L_1$, comparable to the vertical scale height of the atmosphere.
With this assumption, $f(\theta,\phi)$ is no longer singular, as was
the case in Fig. \ref{fmerid}, but instead drops sharply for $\theta
\gtrsim 85^\circ$ and vanishes exactly for $\theta = 90^\circ$ along
the $L_1$ meridian ($\phi=0^\circ$). The resulting $f$ profile does
not depend much on $q$ (the binary mass ratio) and we take here
$q=0.1$.

\subsection{Irradiation}
\label{sectirradiation}

\begin{figure}[t] %  figure placement: here, top, bottom, or page
   \centering
   \includegraphics[width=7cm,angle=90]{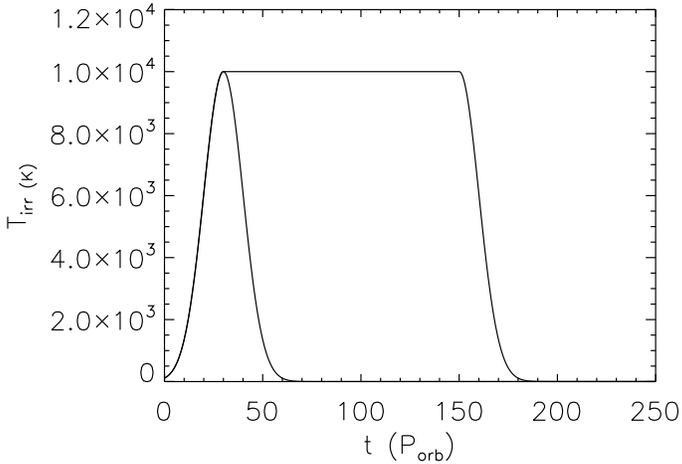}
   \caption{Time profiles of normal outburst and of superoutburst.
   The superoutburst profile has the same rise and
   fall but has a long plateau at maximum.}
   \label{outburst}
\end{figure}

It is still debated if the accretion flux can significantly heat the
secondary atmosphere; we leave this problem for a future
investigation. Here we model the irradiation of the secondary in a
crude, but simple way: we consider that the accretion flux is
completely absorbed and re-radiated as a black body. For
convenience, we define $T_\mathrm{irr}$ as:

\begin{equation}
    T_\mathrm{irr} = (F_\mathrm{irr}(t)/\sigma)^{1/4}
\end{equation}

\noindent where $F_\mathrm{irr}$ is the irradiation flux and
$\sigma$ is the Stefan-Boltzmann constant. $F_\mathrm{irr}$ (and
consequently $T_\mathrm{irr}$) has a time dependence that account
for the temporal profile of the outburst. We denotes by
$T_\mathrm{irr,0}$ the maximum value of $T_\mathrm{irr}$ reached
during an outburst. For simplicity we take $T_\mathrm{irr,0} = 10^4$
K for all models (see Smak 2004a). We use two outburst profiles, a
``short" one lasting $60$ orbital periods to model ``standard" outbursts 
(note that we are not interested in complex behavior such as the outburst bimodality) 
and a ``long" one lasting five times longer in order to model superoutburst of the SU
Uma subtype (model 1), see Fig. \ref{outburst}. Because of the crudeness of the model the
exact shape and duration of our outburst are not of prime importance.

Irradiation produces a radiative isothermal layer whose column
density depends on the irradiation temperature, on top of the
convective envelope. Using simple vertical structure models for the
secondary, we have computed the expected column density
$\Sigma_\mathrm{irr}$ of the isothermal layer when
$T_\mathrm{irr}=10^4$ K. The results are given in Table
\ref{params_phys}. The initial column density is small (a few tens
g.cm$^{-2}$), because the irradiation flux is negligible, and
increases with time during an outburst. For the sake of simplicity,
we do not consider the very initial stages of the outburst, and
assume $\Sigma = \Sigma_\mathrm{irr}(T=10^4$K$)$ as an initial
condition. Since the precise value of $\Sigma$ affects only the
thermal time, we want to obtain the correct order of magnitude of
the cooling time scale for the gas heated at $T_\mathrm{irr}= 10^4$
K.

The irradiation flux at a given point on the secondary surface
enters as a heating term in the energy equation:
\begin{equation}
\label{heating}
F_+ = \sigma \big ( T^4_\star + \psi T_{\mathrm{irr}}^4 \big )
\end{equation}
\noindent where the intrinsic stellar flux is also included. $\psi$
is a space dependent numerical factor accounting for oblique
incidence and shadowing by the accretion disc on the secondary:
\begin{equation}
\psi = (\vec n . \vec u)\times S(\alpha_s)
\end{equation}

\noindent The first effect is accounted for by the $\vec n. \vec u$
term, where $\vec n$ is the local normal to the surface and $\vec u$
is a unit vector pointing towards the primary, assumed to be the
source of irradiation. The second effect is accounted for by the
$S(\alpha_s)$ term, where $\alpha_s$ is the angle between the
orbital plane and the line joining the primary center of mass and
the running point on the secondary:

\begin{equation}
\tan \alpha_s = \frac{z}{\sqrt{(a-x)^2+y^2}}
\end{equation}
\noindent where $(x,y,z)$ are the Cartesian coordinates of the
running point (in the same coordinates system as defined
previously). We then assume that the disc has a fixed opening angle
$\alpha_d$ (see below) and we compute $S(\alpha_s)$ by:

\begin{equation}
S(\alpha_s) = \frac{1}{2}(1+\tanh \frac{\alpha_s - \alpha_d}{\Delta \alpha_d})
\end{equation}
\noindent where $\Delta \alpha_d$ is an angular size controlling the
thickness of the transition between the shadowed part and the
non-shadowed part, taken to be $\Delta \alpha_d = 0.5^\circ$. The
hyperbolic tangent function enables a smooth transition between
$\alpha_s - \alpha_d \ll \Delta \alpha_d$ and $\alpha_s - \alpha_d
\gg \Delta \alpha_d$.

\begin{figure}[t] %  figure placement: here, top, bottom, or page
   \centering
   \includegraphics[width=7cm,angle=90]{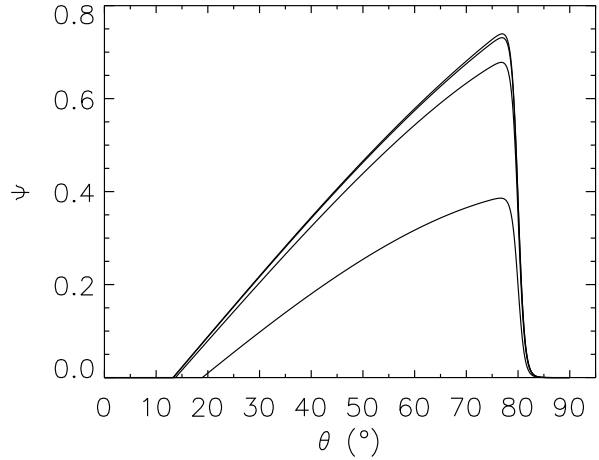}
   \caption{Profile of $\psi$ along different meridians on the secondary:
   $\phi = 0^\circ$ ($L_1$ meridian), $\phi = 5^\circ$, $\phi = 15^\circ$
   and $\phi = 45^\circ$ (from top to bottom). The shadow boundary
   is located at $\theta = 80^\circ$. Note that polar regions are not irradiated.}
   \label{psi}
\end{figure}

We use the numerical code from Hameury et al. (1998) to compute the
vertical structure of the accretion disc during an outburst cycle.
For each model we determine the maximum value of the disc opening
angle $\alpha_d$. %For simplicity we do not take into account the
%temporal variation of the disc opening during an outburst cycle. 
During an outburst, the disc opening is found to be $\alpha_d \sim
4^\circ$ for model 1 and $\alpha_d \sim 6^\circ$ for model 2 and 3.
In both cases this translates into a shadow boundary located at
$\theta \sim 80^\circ$ when measured on the main meridian (the
larger opening of the disc in model 2 and 3 is compensated by larger
secondaries). The quiescent disc has a much lower opening which 
allows for a much smaller shadowed region at the beginning of an outburst. 
We have checked that when the disc enters outbursting state, 
its opening increases very rapidly. Thus it is a good approximation to 
suppose that the disc has its ``outbursting thickness" since the 
very beginning of the outburst.

We use the mapping shown in Fig. \ref{mapping} to project $\psi$ on
the sphere of radius $R_2$ in order to obtain $\psi =
\psi(\theta,\phi)$. Fig. \ref{psi} shows $\psi$ along different
meridians.

\subsection{Equations}

We now turn to the mathematical formulation of the equations. We use
spherical coordinates $(\theta, \phi)$ ($\theta$ is the colatitude
and $\phi$ the longitude) to map the sphere of radius $r_0 = R_2$.
The metric properties of these coordinates are given by the metric
tensor:

\begin{equation}
\label{metric}
(g_{ij}) = \Big (
\begin{array}{cc}
r_0^2 & 0\\
0 & r_0^2\sin^2 \theta\\
\end{array}
\Big )
\end{equation}

\noindent $g^{ij}$ denotes the components of the inverse of matrix
(\ref{metric}) and $g$ denotes its determinant.

We use the following variables: $\Sigma$ is the gas surface density,
$E$, $H = E + P$, and $P$ are the vertically integrated values of
the energy density, enthalpy density, and pressure. $v_\theta$,
$v_\phi$ are the velocity components. For convenience, we introduce
the following notations:

\begin{itemize}
\item ``Barred" quantities denotes a multiplication by  $\sqrt{g}=r_0^2\sin \theta$.
These quantities account for the metric effects; for example $\Sigma dS =
\Sigma \sqrt g d\theta d\phi = \bar \Sigma d\theta d\phi$.
\item ``Tilted" velocity components are angular velocity components:
$\tilde v_\theta = v_\theta/r_0 = \dot{\theta}$ and $\tilde v_\phi =
v_\phi/(r_0\sin \theta) = \dot{\phi}$.
\end{itemize}

With these notations, the Euler equations write:

\begin{eqnarray}
\label{hdeqns}
\frac{\partial}{\partial t} \bar \Sigma  &+&  \frac{\partial }{\partial \theta}\bar
\Sigma \tilde v_\theta +
\frac{\partial }{\partial \phi}\bar \Sigma \tilde v_\phi = 0\\
%---------------
\frac{\partial}{\partial t}  \bar \Sigma \tilde v_\theta &+& \frac{\partial }
{\partial \theta}\bar G^{11} + \frac{\partial }{\partial \phi}\bar G^{12} =
 f\bar\Sigma \tilde v_\phi + \cos \theta \sin \theta \bar G^{22}\\
%---------------
\frac{\partial}{\partial t}  \bar \Sigma \tilde v_\phi &+& \frac{\partial }
{\partial \theta}\bar G^{21} + \frac{\partial }{\partial \phi}\bar G^{22}  =
-f\bar\Sigma \tilde v_\theta - \frac{2}{\tan \theta}\bar G^{12}\\
 %--------------
\frac{\partial}{\partial t} \bar E &+& \frac{\partial }{\partial \theta}
\bar H \tilde v_\theta  + \frac
{\partial }{\partial \phi}\bar H \tilde v_\phi= \bar F_+ - \bar \sigma T^4
\end{eqnarray}

\noindent where $G^{ij} = \Sigma \tilde v^i \tilde v^j + Pg^{ij}$ is
the (symmetric) momentum flux tensor. A radiative cooling term is
included in the r.h.s. of the energy equation in addition to the
heating term (\ref{heating}).

The equations are written in conservative form and differ only
slightly from the Cartesian case, with the inclusion of geometric
source terms in the momentum equations (in addition to the Coriolis
term) and with conservative variables $(\bar \Sigma, \bar \Sigma
\tilde v_\theta, \bar \Sigma \tilde v_\phi, \bar E)$ that are not
physical quantities. We also assume a perfect gas equation of state:

\begin{equation}
P = \Sigma R_g T\mathrm{\ and\  } E = \frac{1}{2}\Sigma v^2 + \frac{1}{\gamma-1}P
\end{equation}

\noindent with an adiabatic index $\gamma = 5/3$ and $R_g = R/\mu$,
$R$ being the perfect gas constant.

Note that we do not include viscosity. As in many astrophysical
processes, the molecular viscosity is negligible but turbulent
viscosity could be significant. In a geostrophic state, Osaki \&
Meyer (2003) showed that viscosity is responsible for a drift across
isobars, but the expected value of turbulent viscosity yields only a
very small drift velocity.

\subsection{Simulation setup}

For each simulation, we start with an isothermal envelope with
temperature $T_\star$ and surface density $\Sigma_\mathrm{irr}$ (see
Table \ref{params_phys}). The envelope is initially at rest,
\emph{i.e.} we neglect the quiescence steady surface flow feeding
mass transfer. The order of magnitude of the quiescent flow speed is
$\sim 0.01c_\mathrm{s}$ (see Lubow \& Shu 1975), small compared to
surface velocities found in outburst (see Sect. \ref{results}).
Irradiation is then turned on with the temporal profile simulating
an outburst or a superoutburst. Our simulations span a time length
of $240\ P_\mathrm{orb}$.

\section{The numerical code}
\label{code}

The Cartesian-like form of our equations enables us to apply
directly any scheme developed for Cartesian coordinates. Here we use
the TVD-MacCormack scheme (see Yee 1987) to solve equations $(8-11)$
for $(\bar \Sigma, \bar \Sigma \tilde v_\theta, \bar \Sigma \tilde
v_\phi, \bar E)$. This scheme is a finite difference
predictor-corrector scheme followed by a TVD step. The ``Total
Variation Diminishing" step ensures that non physical oscillations
do not appear (see Hirsch 1990). The two step nature of the scheme
enables second order accuracy in time as well as a second order
discretization of the source terms. To obtain second order in space,
the spatial discretization alternates between backward and forward
differencing during the predictor and corrector steps. To avoid
error accumulation, the order of forward/backward differencing is
switched at each time-step. In steep gradients region, the TVD step
adds a diffusive flux locally altering the scheme order. The $\sin
\theta \cos \theta$ term in equation (6) is discretized in such a
way that the physical equilibrium ($P$ = cst.) is also a solution of
the discretized equation.

Due to the explicit character of the scheme, the numerical stability
of the scheme is subject to the well known Courant-Friedrichs-Levy
(CFL) condition; the time step is restricted by (see Hirsch 1990):

\begin{equation}
\Delta t < \Delta t_\mathrm{max} = \frac{1}{\tilde c^\mathrm{max}_\theta/
\Delta \theta + \tilde c^\mathrm{max}_\phi/\Delta \phi}
\end{equation}

\noindent where $\tilde c^\mathrm{max}_\theta = \mathrm{max \big
(}c_\mathrm{s}/r_0 + | \tilde v_\theta | \mathrm{\big)}$ and $\tilde
c^\mathrm{max}_\phi = \mathrm{max \big(}c_\mathrm{s}/(r_0\sin
\theta) + | \tilde v_\phi | \mathrm{\big)}$. In our simulations we
take $\Delta t = \lambda \Delta t_\mathrm{max}$ with $\lambda =
0.95$ (a simulation with $\lambda = 0.4$ has been checked to give
identical results).

\begin{figure*}[t] %  figure placement: here, top, bottom, or page
\parbox{\linewidth}{\center 2500 K \includegraphics[scale=0.25,angle=180]
{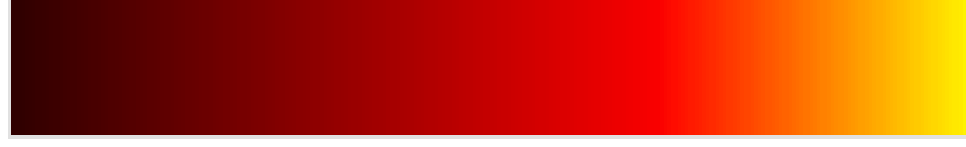} 10000 K}

\center $t=0.7\ P_\mathrm{orb}$\\
\parbox{\linewidth}{
\parbox{0.33\linewidth}{\includegraphics[width=\linewidth]{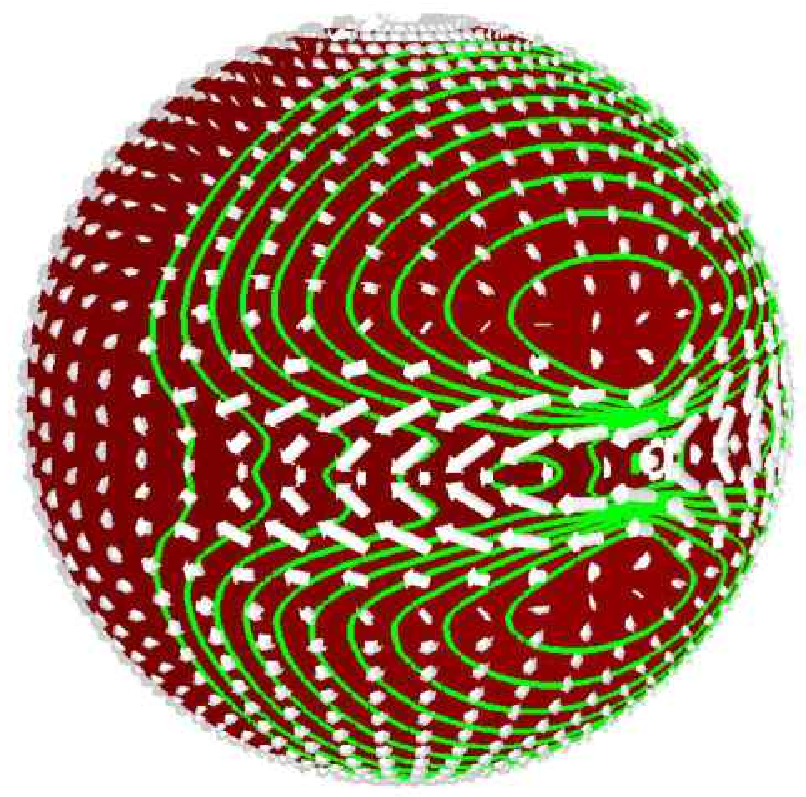}}
\parbox{0.33\linewidth}{\includegraphics[width=\linewidth]{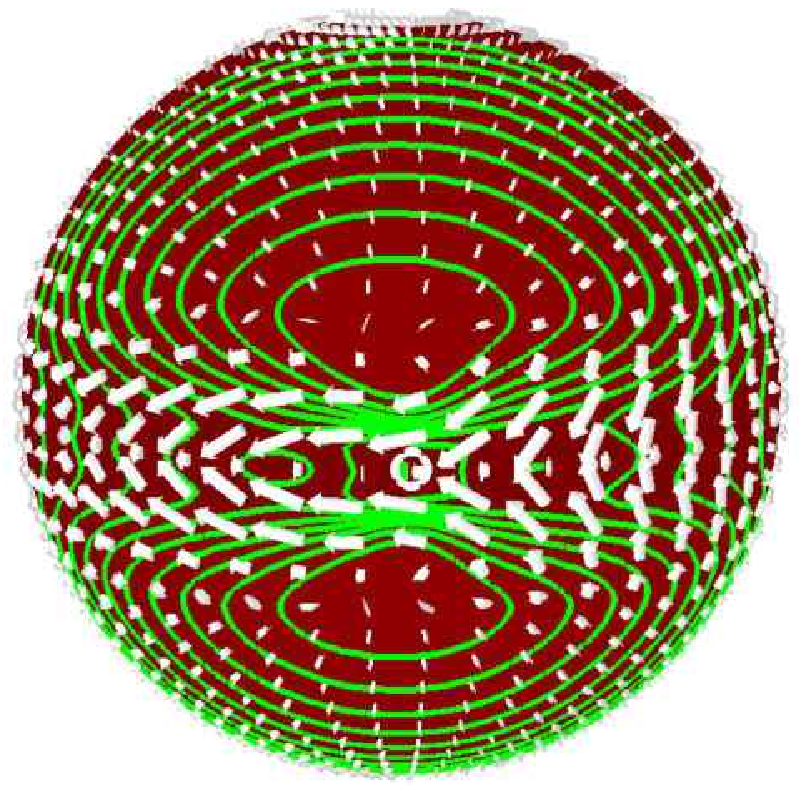}}
\parbox{0.33\linewidth}{\includegraphics[width=\linewidth]{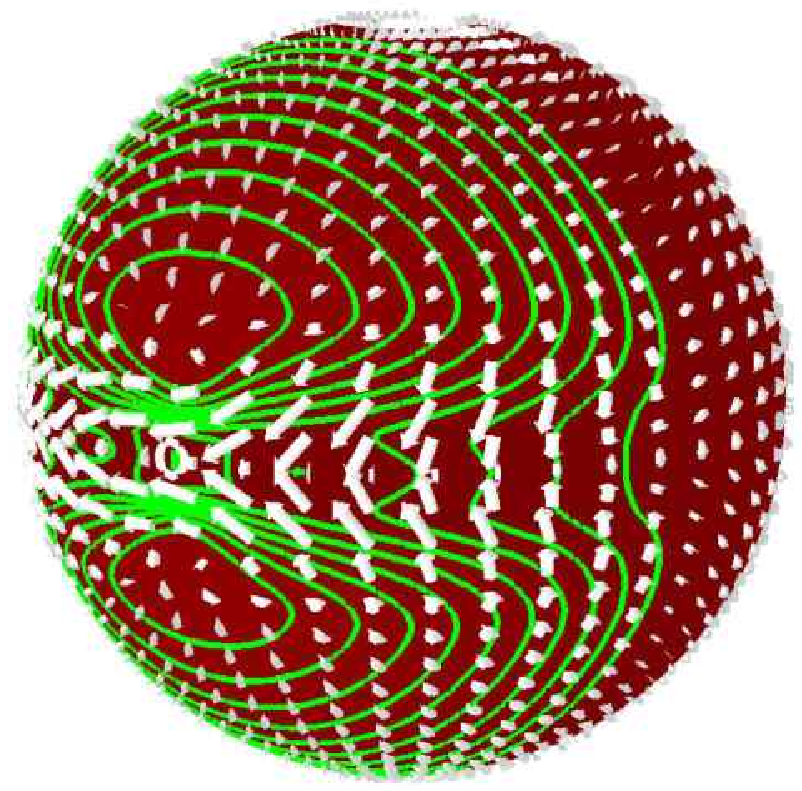}}
}\\

\center $t=2.5\ P_\mathrm{orb}$\\
\parbox{\linewidth}{
\parbox{0.33\linewidth}{\includegraphics[width=\linewidth]{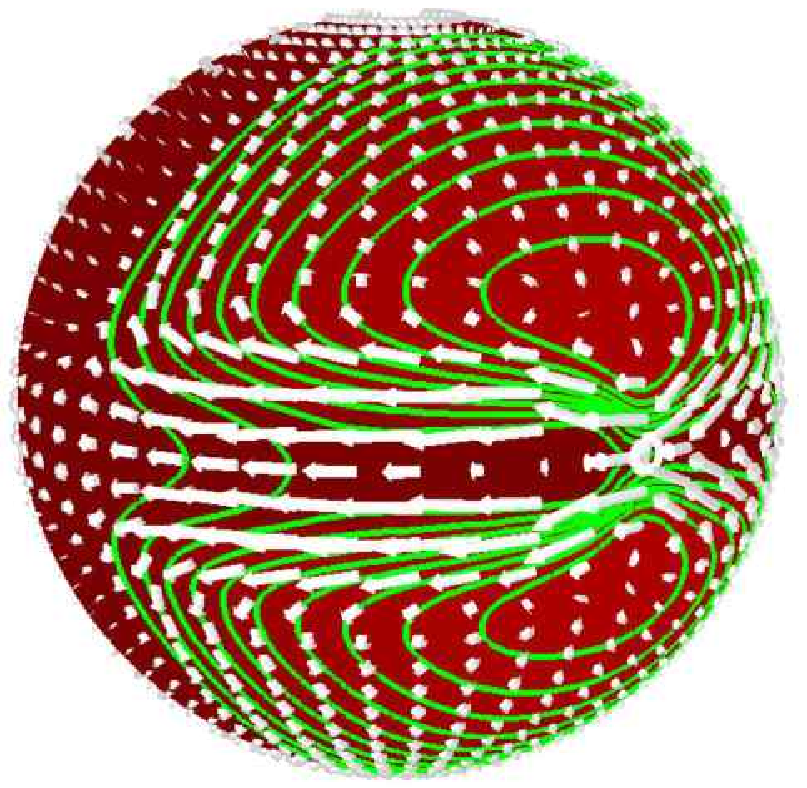}}
\parbox{0.33\linewidth}{\includegraphics[width=\linewidth]{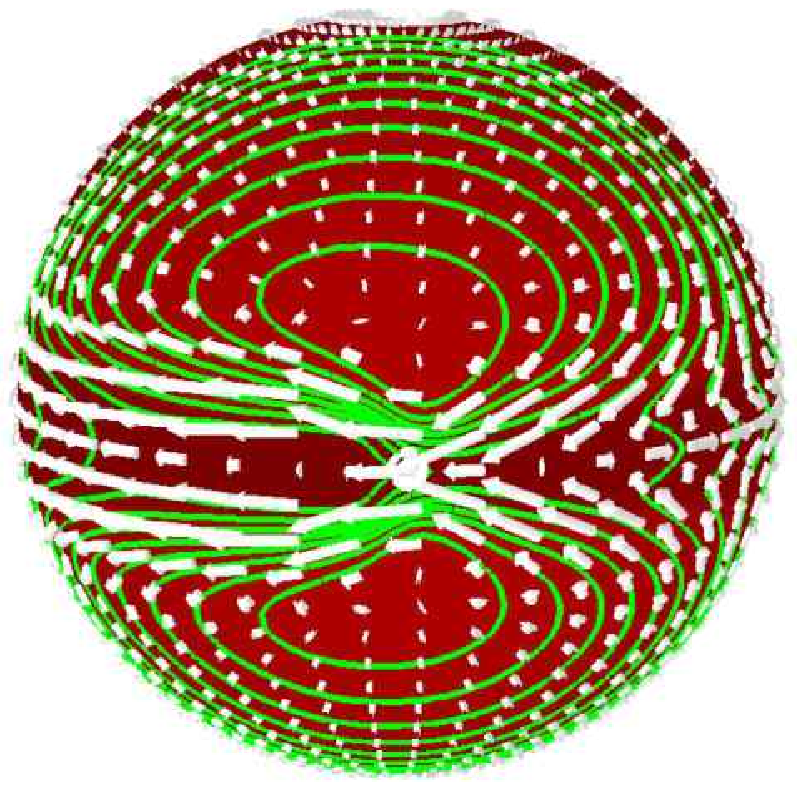}}
\parbox{0.33\linewidth}{\includegraphics[width=\linewidth]{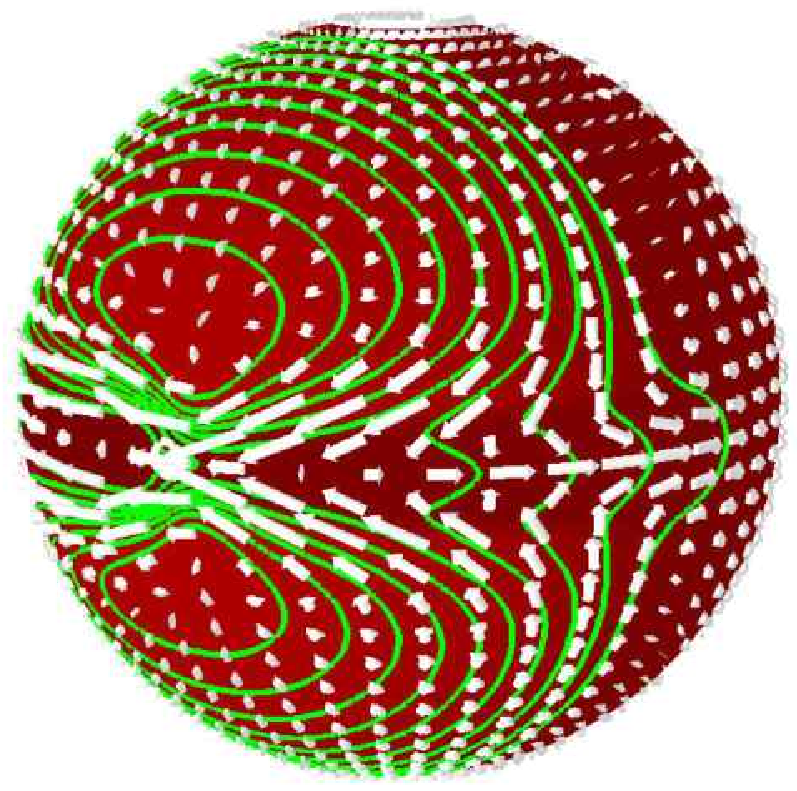}}
}\\

\center $t=5\ P_\mathrm{orb}$\\
\parbox{\linewidth}{
\parbox{0.33\linewidth}{\includegraphics[width=\linewidth]{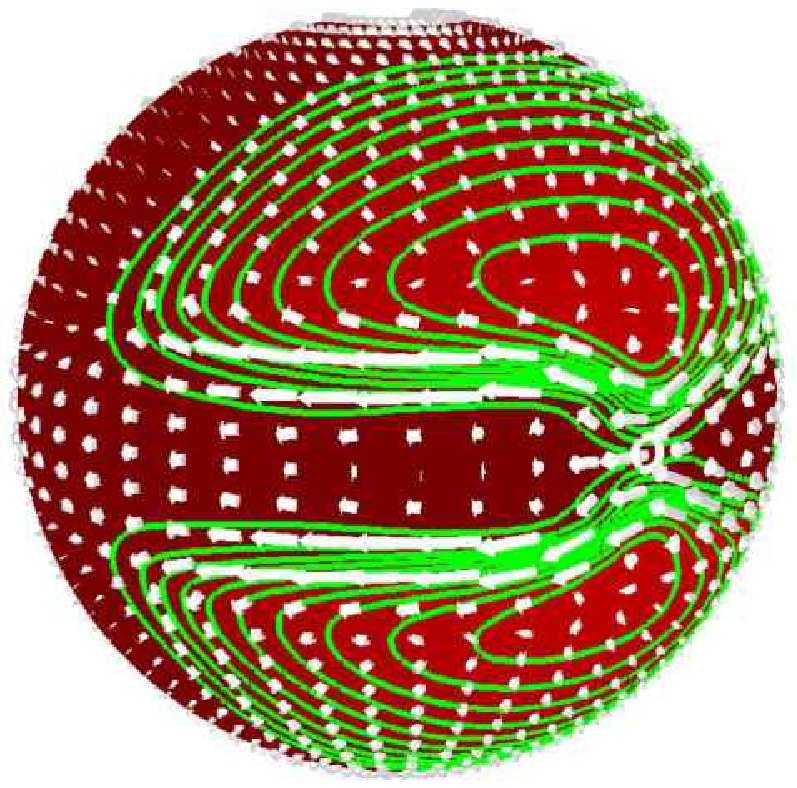}}
\parbox{0.33\linewidth}{\includegraphics[width=\linewidth]{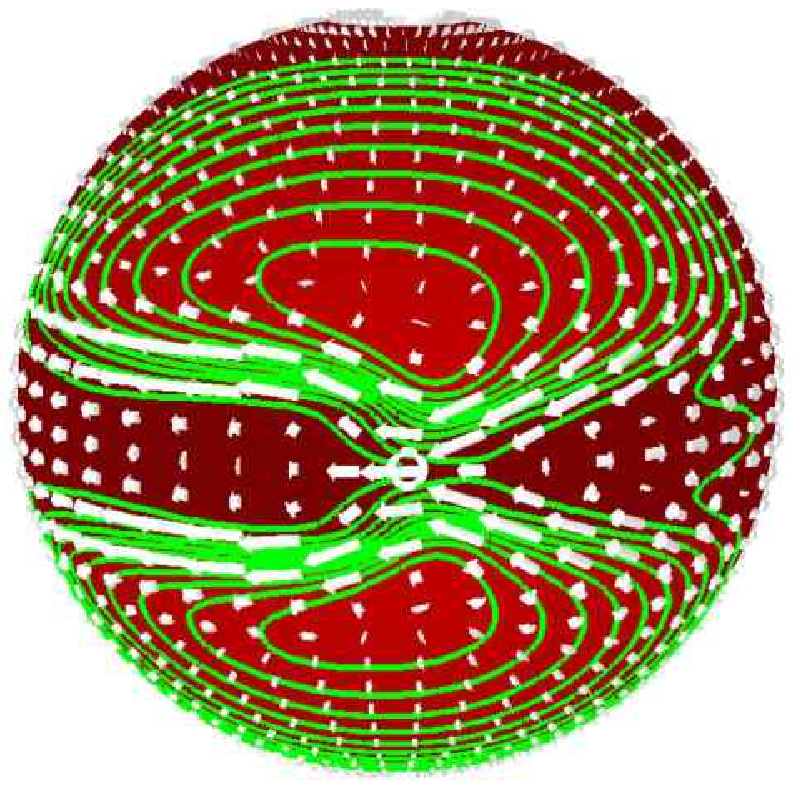}}
\parbox{0.33\linewidth}{\includegraphics[width=\linewidth]{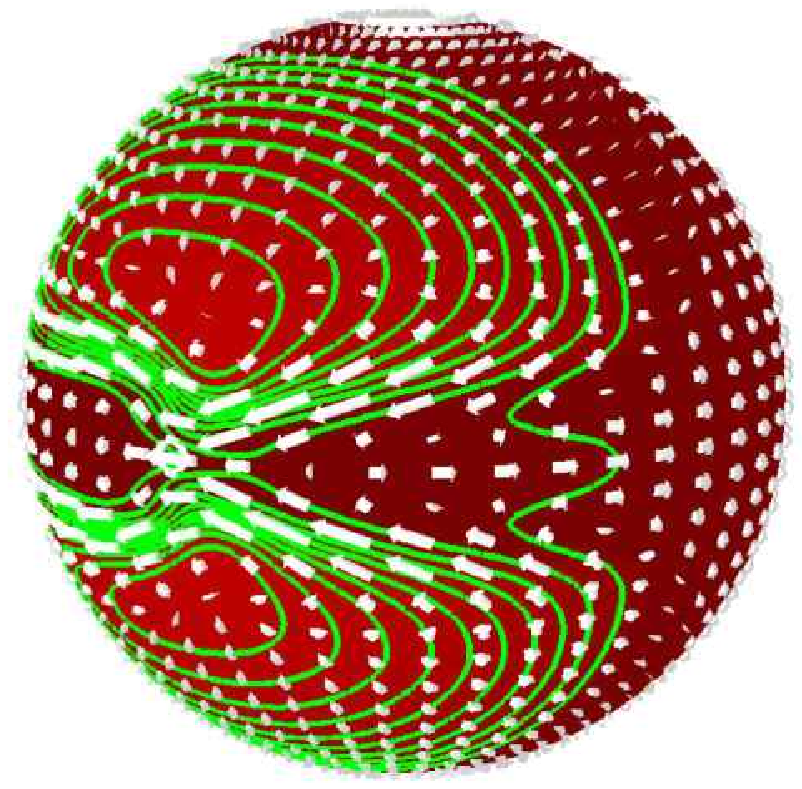}}}
\\

   \caption{Snapshots at $t=0.7, 2.5, 5\ P_\mathrm{orb}$. Each snapshot shows
   the temperature field in color gradient (see the color legend), isobars and
   the velocity field. The central panel shows the secondary face on; in the left
   and right panels, it is rotated by +/- $30$ degrees. The $L_1$ region is
   marked by a circle whose surface is equal to the stream cross section
   (see Eq. (17)). For the sake of figure readability, the magnitude of the velocity field has been multiplied by 3 at $t = 0.7\ P_\mathrm{orb}$ and by 2 at $t = 2.5\ P_\mathrm{orb}$.}
   \label{sim1}
\end{figure*}

\begin{figure*}[t] %  figure placement: here, top, bottom, or page
\parbox{\linewidth}{\center 2500 K \includegraphics[scale=0.25,angle=180]
{colormap.eps} 10000 K}

\center $t=18\ P_\mathrm{orb}$\\
\parbox{\linewidth}{
\parbox{0.33\linewidth}{\includegraphics[width=\linewidth]{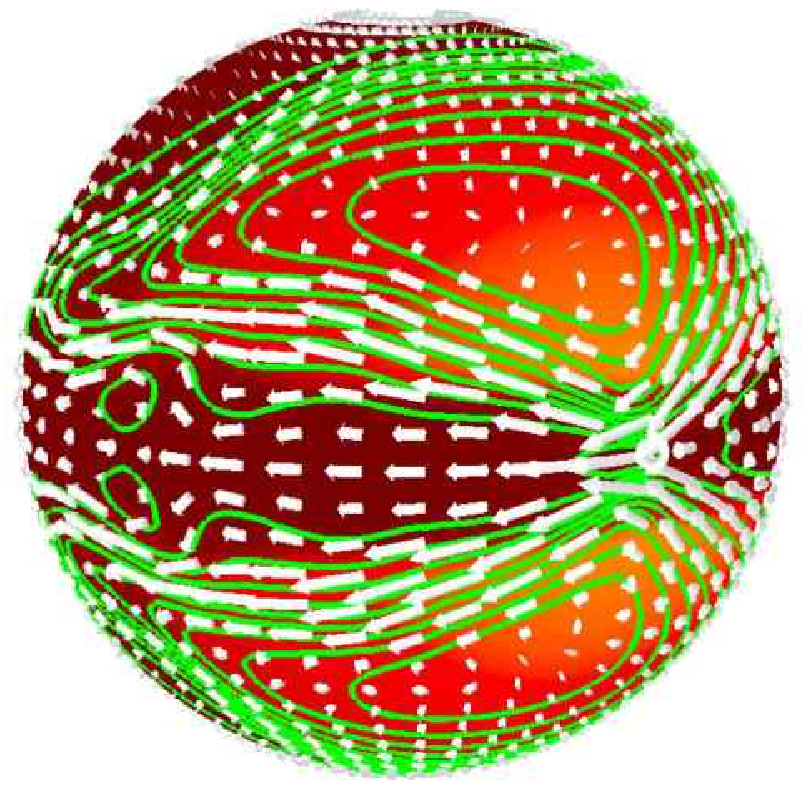}}
\parbox{0.33\linewidth}{\includegraphics[width=\linewidth]{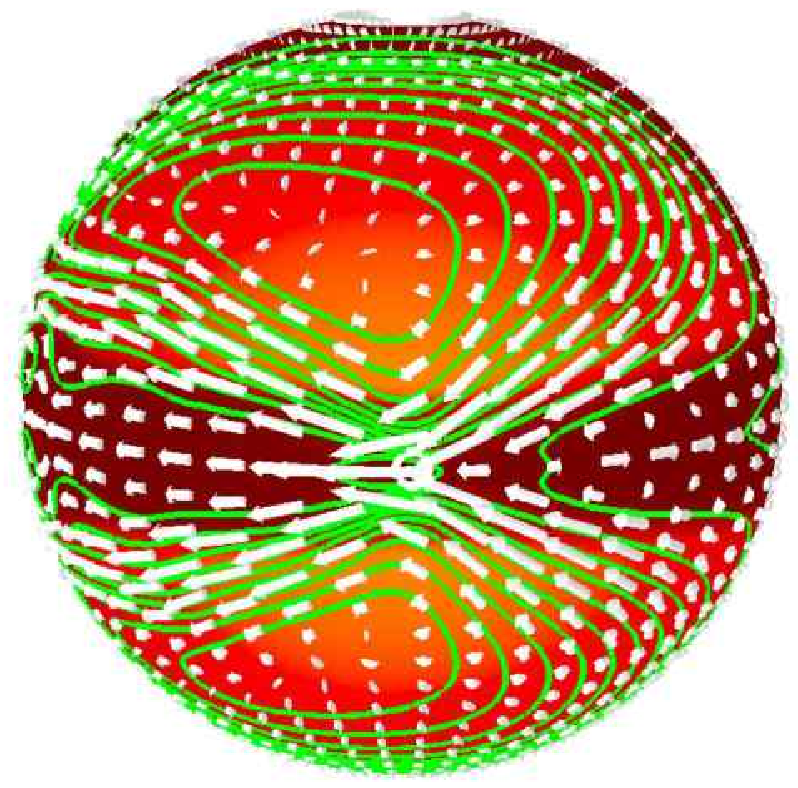}}
\parbox{0.33\linewidth}{\includegraphics[width=\linewidth]{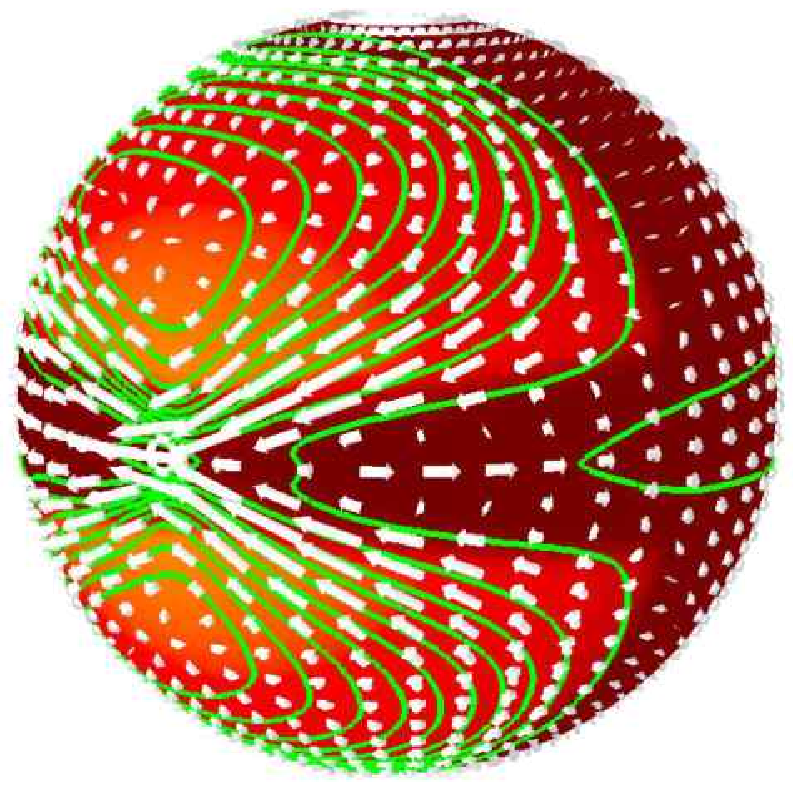}}}

\center $t=30\ P_\mathrm{orb}$\\
\parbox{\linewidth}{
\parbox{0.33\linewidth}{\includegraphics[width=\linewidth]{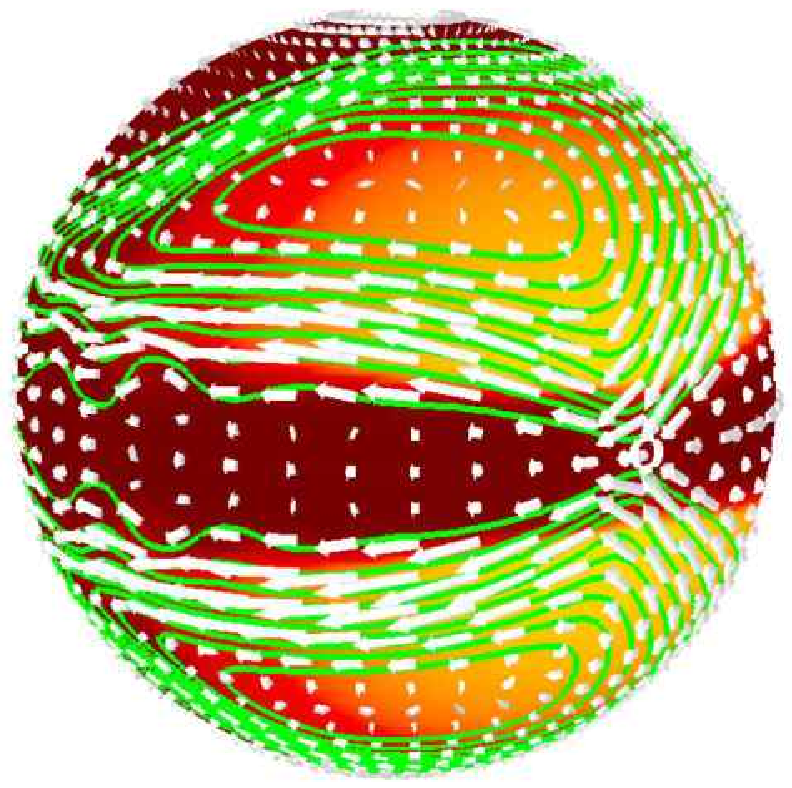}}
\parbox{0.33\linewidth}{\includegraphics[width=\linewidth]{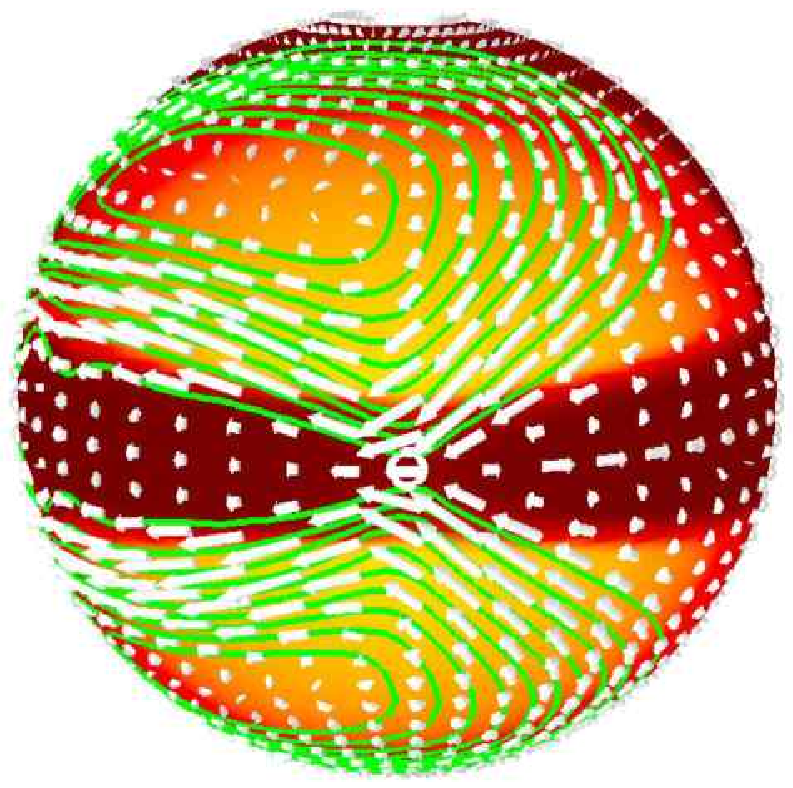}}
\parbox{0.33\linewidth}{\includegraphics[width=\linewidth]{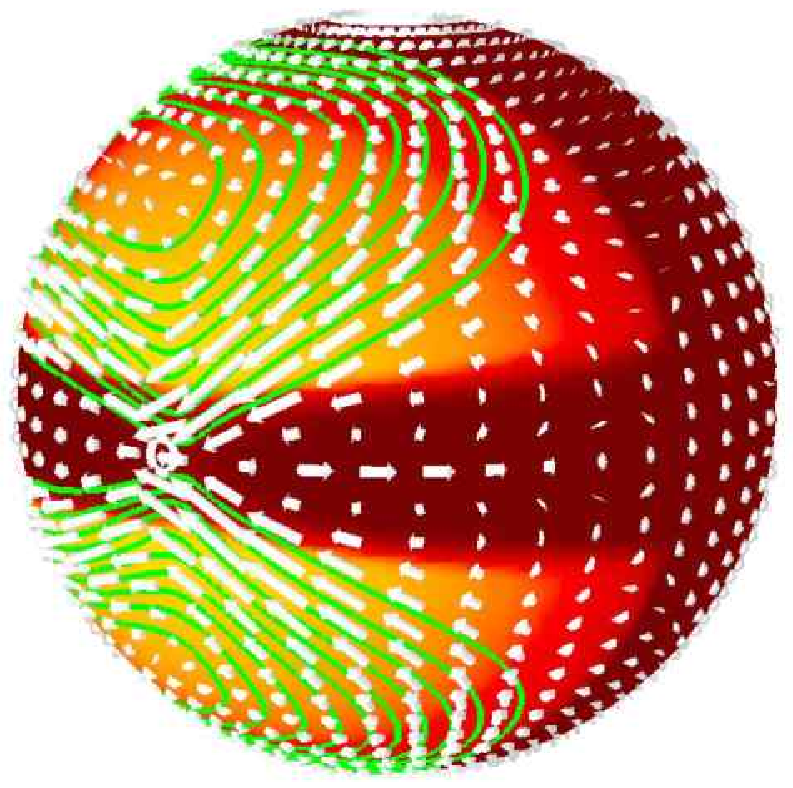}}}

\center $t=45\ P_\mathrm{orb}$\\
\parbox{\linewidth}{
\parbox{0.33\linewidth}{\includegraphics[width=\linewidth]{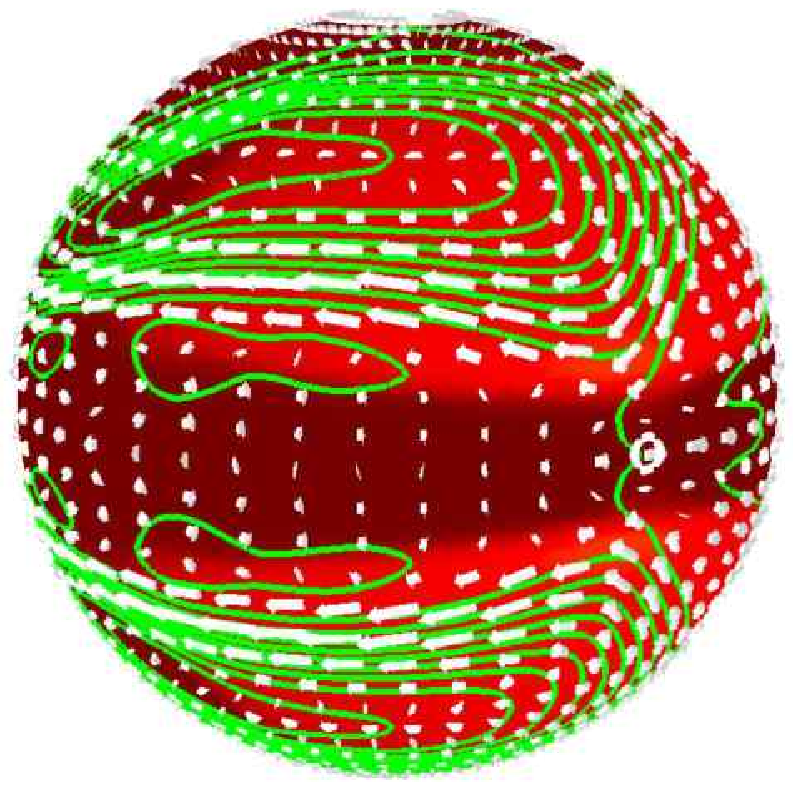}}
\parbox{0.33\linewidth}{\includegraphics[width=\linewidth]{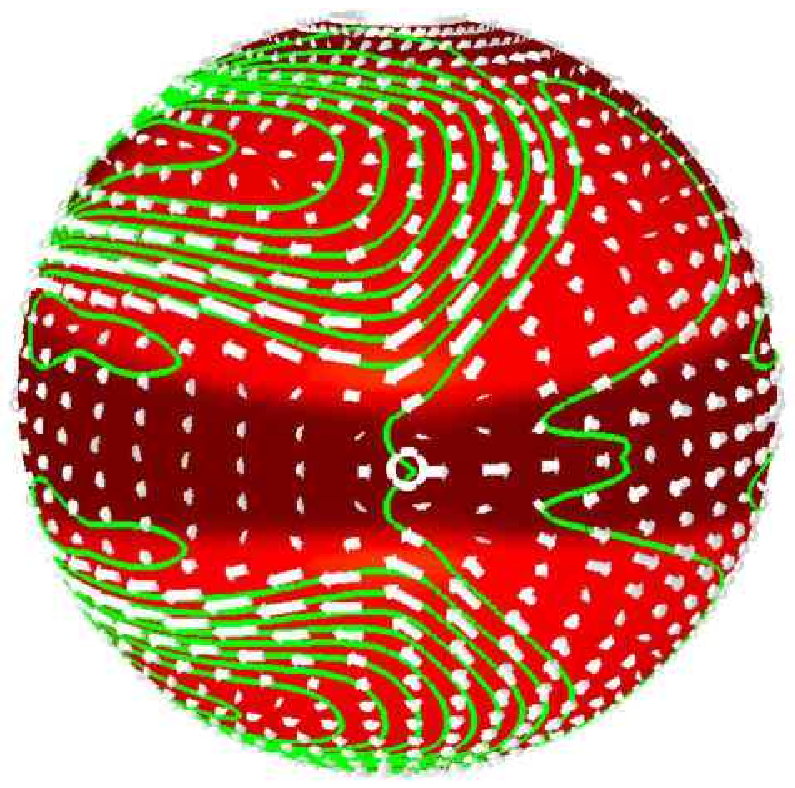}}
\parbox{0.33\linewidth}{\includegraphics[width=\linewidth]{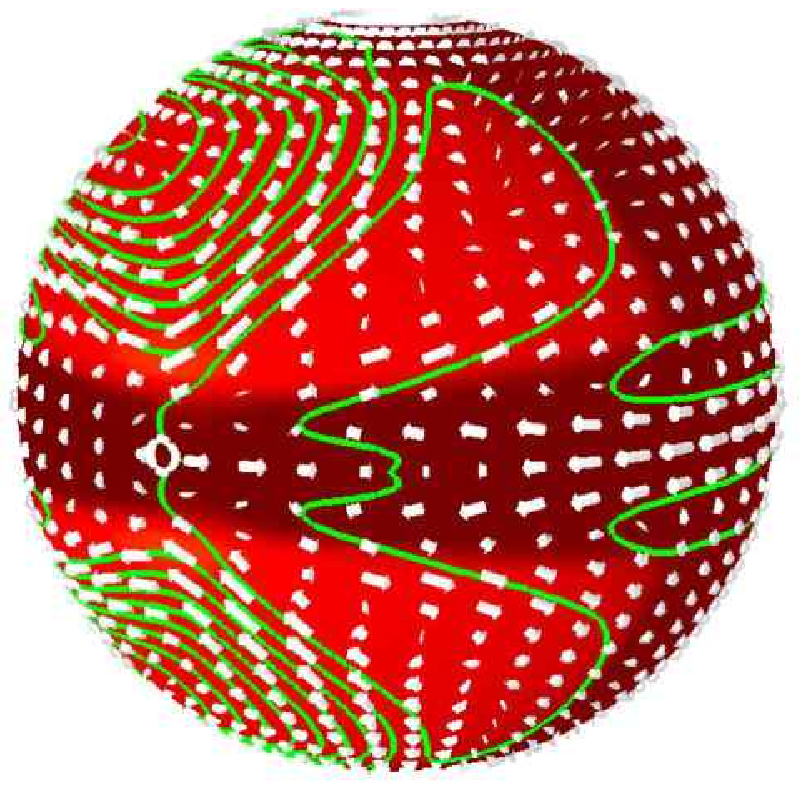}}}

\center $t=45\ P_\mathrm{orb}$ during a superoutburst\\
\parbox{\linewidth}{
\parbox{0.33\linewidth}{\includegraphics[width=\linewidth]{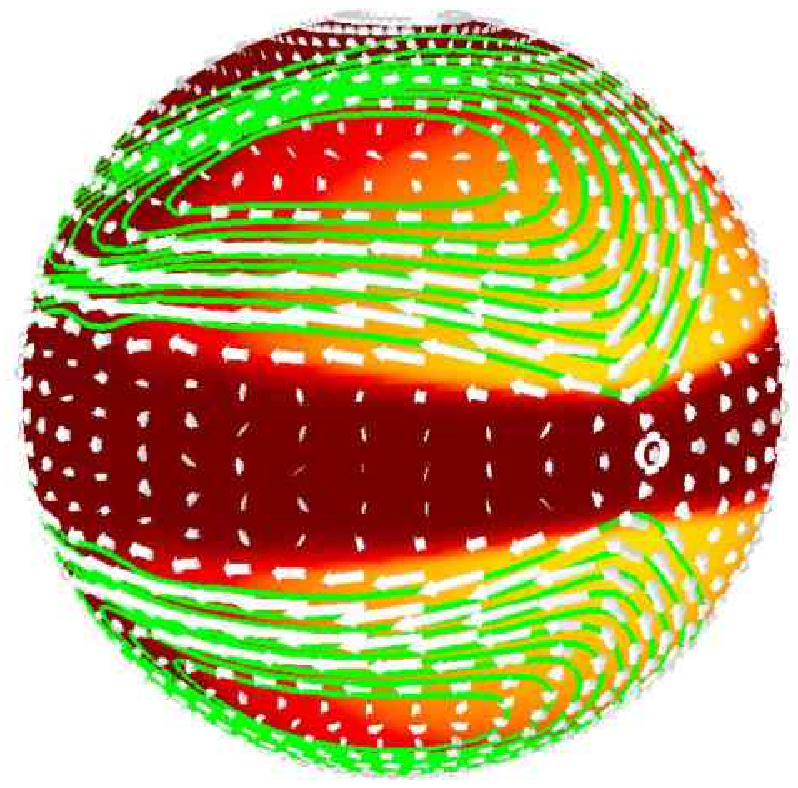}}
\parbox{0.33\linewidth}{\includegraphics[width=\linewidth]{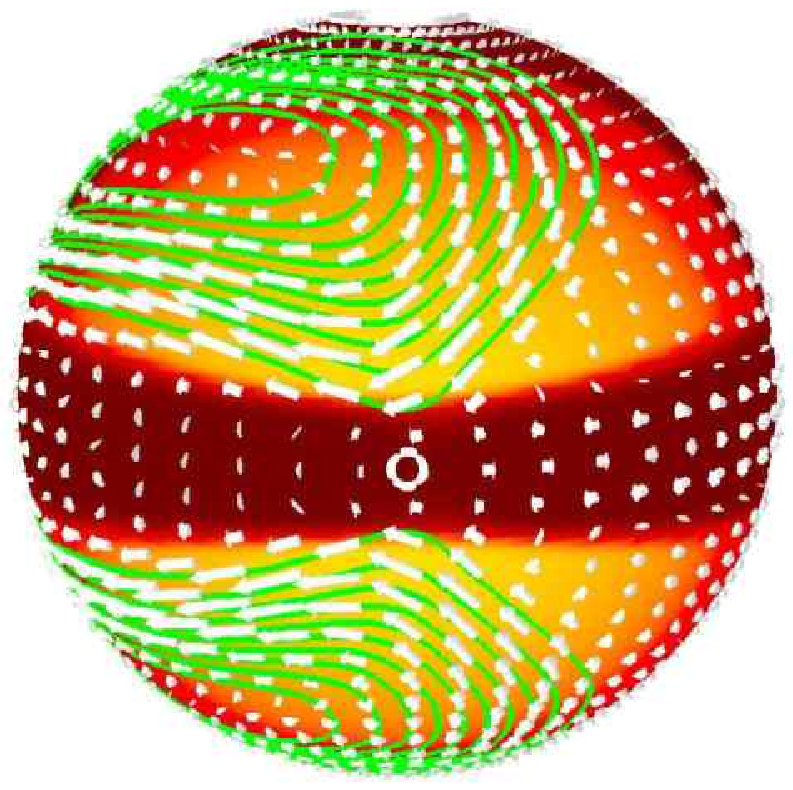}}
\parbox{0.33\linewidth}{\includegraphics[width=\linewidth]{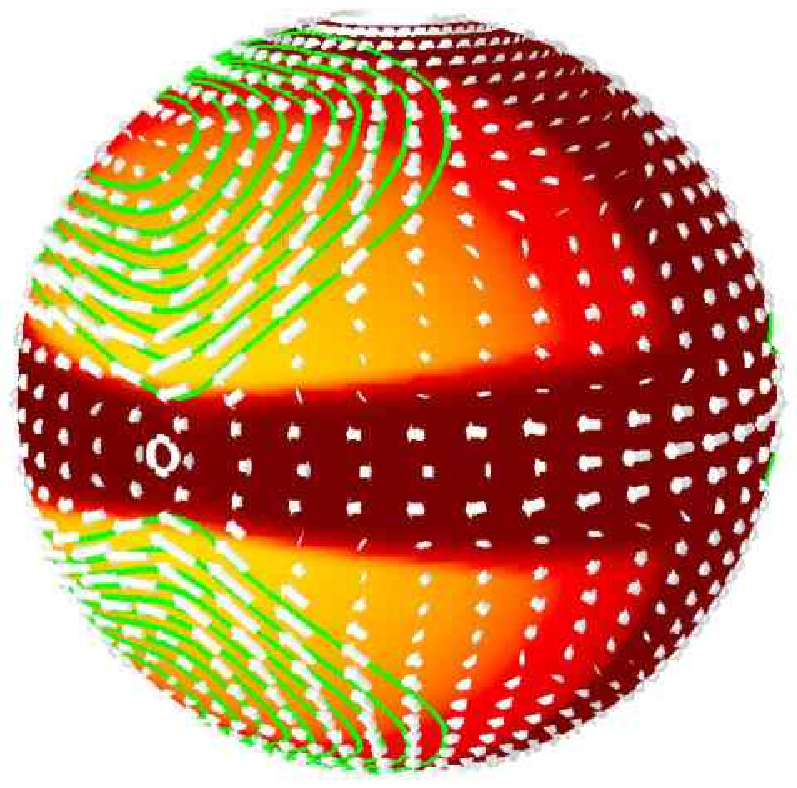}}}

   \caption{Snapshots at $t=18, 30, 45\ P_\mathrm{orb}$.  Each snapshot shows the
   temperature field in color gradient (see the color legend), isobars and the velocity field}
   \label{sim2}
\end{figure*}

We consider a physical domain $\theta_0 < \theta < \pi-\theta_0$ and
$ -\pi < \phi< \pi$. We therefore exclude the two polar caps in
order to avoid difficulties around the poles that are singular
points in spherical coordinates. This is not a serious problem, as
the poles are not irradiated (see Fig. \ref{psi}). The grid point
accumulation occurring there would severely restrict the CFL
condition. In the following, we choose $\theta_0 = 10^\circ$ and we
have checked that this particular choice does not influence our
results. We use an uniform grid with $151\times200$ grid points and
we have checked the effect of numerical resolution for model 1 on a
$76\times100$ and a $301\times400$ grid. As expected the low
resolution run shows signs of enhanced numerical dissipation and the
high resolution run shows a more turbulent behavior as compared to
our standard resolution, but the main features and results presented
here are not significantly modified by resolution effects (changes
are less than $10\%$). Note that the number of grid points in
$\theta$ is odd in order to have a grid point at the equator.
Periodic boundary conditions are used at $\phi = \pm \pi$ and free
outflows conditions are considered at $\theta_0$ and $\pi -
\theta_0$ (we checked that this does not affect our results).

\section{Results}
\label{results}

In this section, the velocity, density and temperature at $L_1$
refer to average quantities computed using a gaussian kernel
centered on $(\theta=90^\circ,\phi=0^\circ)$ with a FWHM equal to
the radius of the cross section of the stream leaving the secondary
(see Eq. (\ref{cross_section}) below).

\begin{figure*}[t] %  figure placement: here, top, bottom, or page
\parbox{0.5\linewidth}{\includegraphics[width=7cm,angle=90]{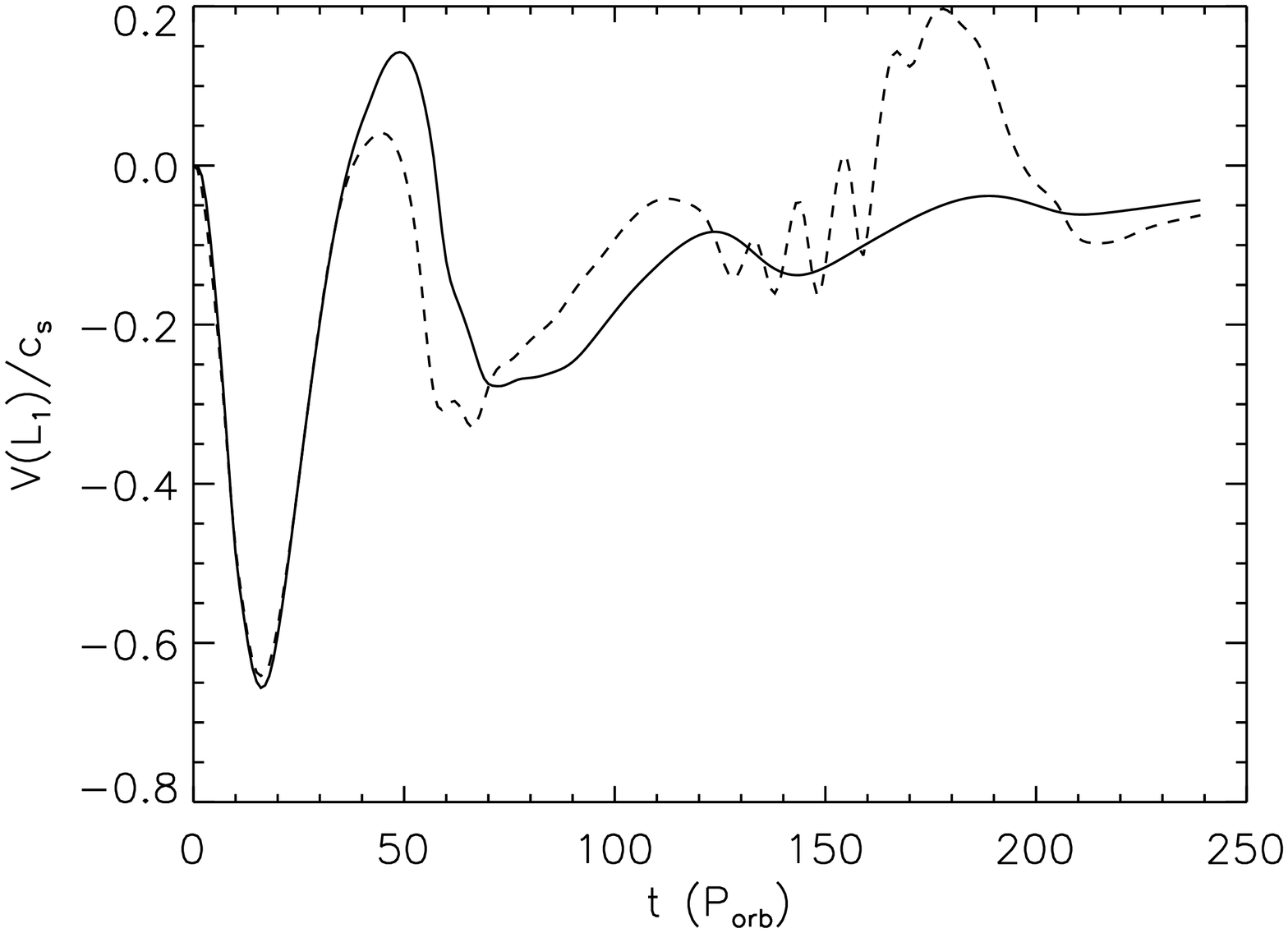}}
\parbox{0.5\linewidth}{\includegraphics[width=7cm,angle=90]{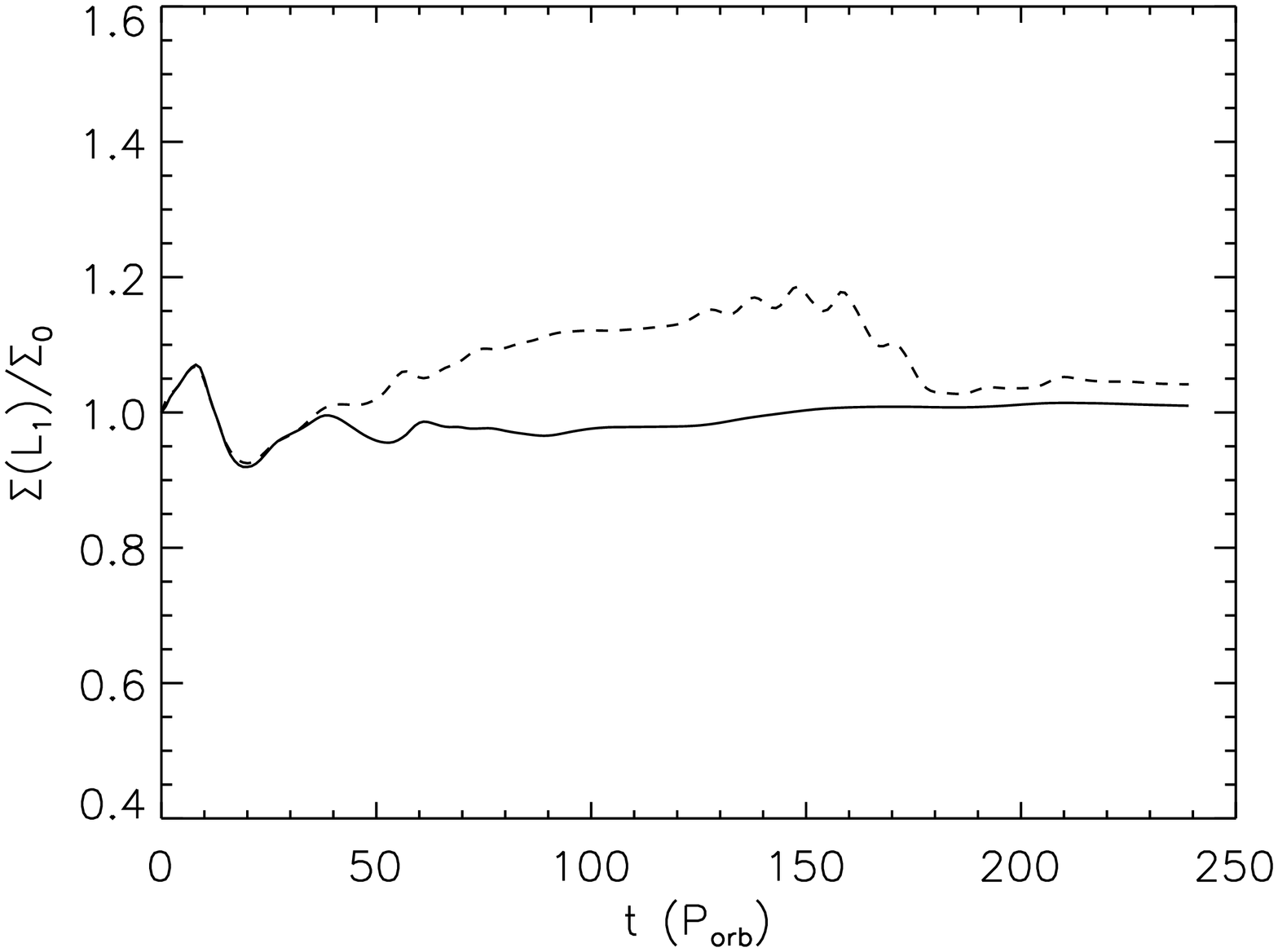}}
   \caption{Left panel: Evolution of the velocity at $L_1$ normalized by the
   isothermal sound speed (computed with the temperature at $L_1$) during an outburst
   (solid line) and a superoutburst (dashed line). The maximum velocity is reached
   at $t=18\ P_\mathrm{orb}$ (see Fig. \ref{sim2}). Right panel: Evolution of
   the surface density at the $L_1$ point normalized by its initial value
   surface density during an outburst (solid line) and a superoutburst (dashed line).}
   \label{VL1}
\end{figure*}

\subsection{Model 1: the SU Uma class}

We first discuss normal outbursts. As expected, irradiation
increases the surface temperature of the secondary and, as a result,
pressure gradients form (see Fig. \ref{sim1}). In the presence of
pressure gradients, the flow tends to reach geostrophism where the
Coriolis force balances pressure gradients (i.e. velocity is along
constant pressure line). This dynamical phenomenon is known as
``Rossby adjustment" or ``geostrophic adjustment" (see Pedlosky 2003).
Initially, vertical pressure gradients drive a vertical acceleration
of the fluid, pushing it towards the equator. However the Coriolis
force quickly deflects the flow westwards and this effect is
stronger in the vicinity of $L_1$. This leads to an oscillatory
pattern of the velocity field (see snapshots at $t=0.7\
P_\mathrm{orb}$ in Fig. \ref{sim1}) along the shadow boundary of the
disc. This pattern drives the flow toward $L_1$. In addition,
vertical pressure gradients appearing at high longitude, where the
normal irradiation flux drops, are also adjusted and as a result, a
clockwise circulation flow sets in. At $t=2.5\ P_\mathrm{orb}$ (see
Fig. \ref{sim1}), the flow has reached geostrophism. The flow
crosses isobars only in the close vicinity of $L_1$, as the Coriolis
force is weaker there (see Sect. \ref{geom_cons}) and the flow
speed larger. Note that the geostrophic state is significantly
different from the one expected from the initial pressure gradients;
this shows the complexity of the geostrophic adjustment process
during which both the flow velocity and the pressure gradients vary,
the irradiation flux being the only fixed quantity.

The overall situation can be depicted as follows: irradiation drives
a giant anti-cyclonic perturbation on each hemisphere of the
secondary. Furthermore, the associated clockwise circulation goes
through the $L_1$ point. In the same time, an equatorial jet
traveling westwards forms at longitude $\phi \sim -60^\circ$ as well
as a weaker eastwards one traveling one at $\phi \sim 60^\circ$ (see
snapshots at $t=2.5\ P_\mathrm{orb}$ in Fig. \ref{sim1}).

As time goes on, pressure gradients increase with irradiation and,
as a result, the circulation strengthens (see Fig. \ref{sim2} at $t
= 30\ P_\mathrm{orb}$, the outburst maximum). The left panel of Fig.
\ref{VL1} shows the magnitude of the velocity at $L_1$. The maximum
velocity is reached at $t = 18\ P_\mathrm{orb}$ where $v_{\phi} \sim
0.6c_\mathrm{s}$ (for symmetry reason $v_{\theta} = 0$). The
corresponding snapshot is shown in Fig. \ref{sim1}. The right panel
of Fig. \ref{VL1} shows the evolution of the surface density at
$L_1$. In the early stage of the outburst, the surface density
increases by a few percent as matter is initially driven toward the
equator. Later on, it decreases temporarily by a few percents due to
the strong surface flow through $L_1$ and as a consequence of mass
conservation.

Snapshots at different times (see Figs. \ref{sim1}, \ref{sim2}) show
that the anti-cyclonic perturbation slowly drifts westward. This is
a natural consequence of its clockwise circulation and this shows
that no steady state can ever be reached. As a consequence, the
circulation flow initially transiting by the $L_1$ point moves away
from $L_1$ (see left panel of Fig. \ref{VL1} for $t < 50\
P_\mathrm{orb}$). Note also that the temperature distribution at the
surface of the secondary is not much affected by the flow, as would
be the case if heat transport by advection were important. This is
an important point that will be discussed in more details later on.

The subsequent increase of the velocity at $L_1$ in Fig. \ref{VL1}
(left panel, at $t\sim 70\ P_\mathrm{orb}$) is due to the transit of
the westwards jet.

At the outburst end, the residual circulation flow slowly decays
(due to numerical dissipation) on a time scale shorter than the time
interval between outbursts; physical viscosity (e.g. due to
turbulence) would bring the velocity field to zero before the next
outburst starts.

\begin{figure*}[t] %  figure placement: here, top, bottom, or page
\parbox{0.5\linewidth}{\includegraphics[width=0.8\linewidth,angle=90]{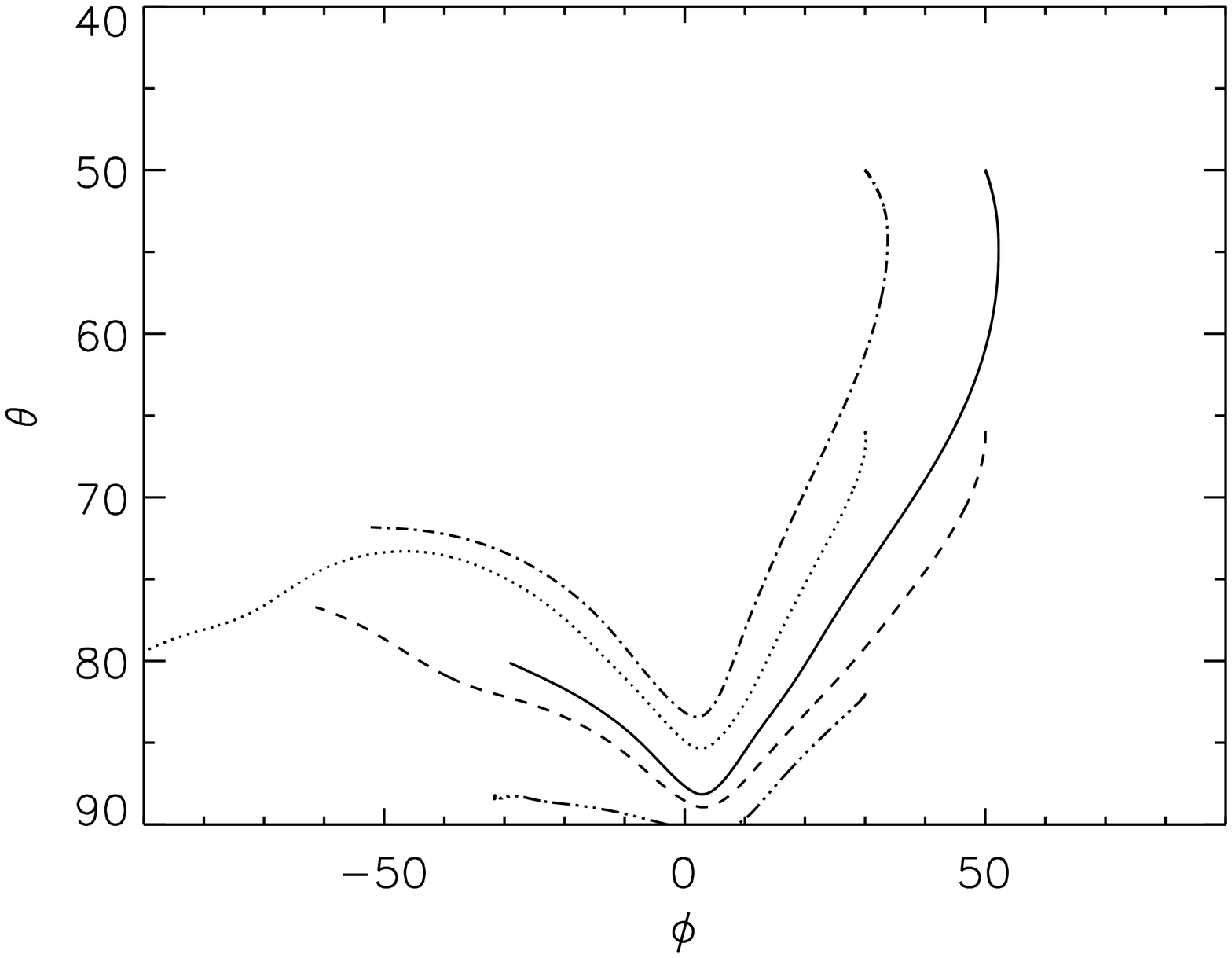}}
\parbox{0.5\linewidth}{\includegraphics[width=0.8\linewidth,angle=90]{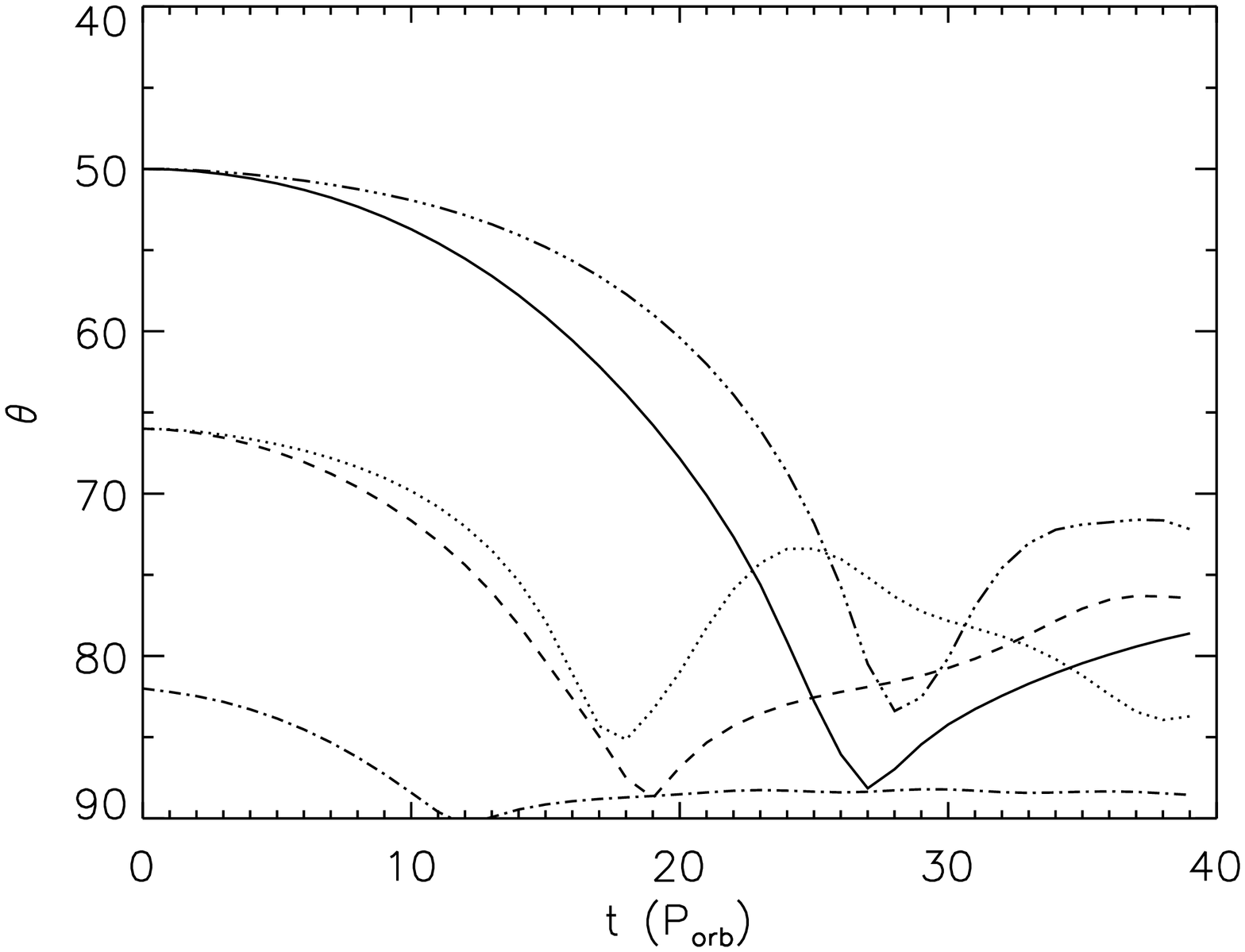}}
   \caption{Left panel: Particles pathlines from $t=0$ to $t=40\ P_\mathrm{orb}$.
   Right panel: Time evolution of the colatitude of each particles. The shadow boundary
   is at $\theta = 80^\circ$.}
   \label{parts}
\end{figure*}

Numerical simulations of a superoutburst shows the same features
during the initial phase, as expected. Even after $t=30\
P_\mathrm{orb}$, when a normal outburst would have started to
decline, the surface flow is not very different in the normal and
superoutburst cases, see Fig. \ref{sim2}. Due to the maintained
irradiation, the strength of the surface flow does not decay as it
does in the normal outburst case. However, the anti-cyclonic
perturbation drifts westward in the same way as it does in the
normal outburst case, also suppressing the flow transiting by $L_1$
(see left panel of Fig. \ref{VL1}). The subsequent evolution shows a
turbulent like behavior that slowly decays when the superoutburst
ends. The surface density at $L_1$ (see right panel of Fig.
\ref{VL1} is increased by at most $20 \%$ during a superoutburst.

\begin{figure}[t] %  figure placement: here, top, bottom, or page
   \centering
   \includegraphics[width=7cm,angle=90]{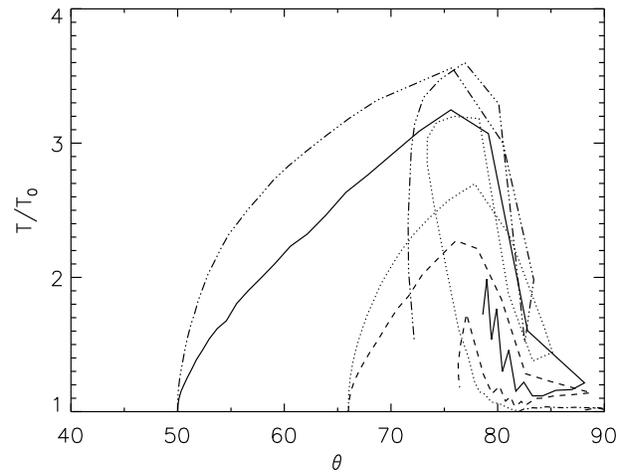}
   \caption{Temperature vs colatitude for each particle ($T_0=2500$ K,
   the shadow boundary is at $\theta = 80^\circ$). Particles cool very
   rapidly as they enter the shaded region.}
   \label{Tparts}
\end{figure}

In order to compare our results with those from Smak (2004a), we now
follow the trajectories of test particles distributed in the
atmosphere and initially at rest. Fig. \ref{parts} (left panel)
shows a few example of such trajectories. Each trajectory lasts for
$40$ orbital periods. This figure shows that particles initially at
longitude $\phi \sim40^\circ - 50^\circ$ reach the close vicinity of
$L_1$. The arrival time in the $L_1$ region can be obtained from the
right panel of Fig. \ref{parts} which shows the time evolution of
each particle colatitude. Fig. \ref{Tparts} shows the temperature of
each particle vs colatitude. The particles are heated during their
trajectory towards $L_1$ and they cool down very quickly once they
enter the shadowed region. As a consequence, the temperature of the
$L_1$ point remains very close to its quiescent value ($2500$ K),
contrary to Smak (2004a). The reason for this is that the cooling
time scale is shorter than the transit time towards $L_1$: as shown
in the right panel of Fig. \ref{parts}, the time scale for crossing
the shadow boundary is of the order of few orbital periods. The
cooling time scale is:

\begin{equation}
\tau_\mathrm{cool} = \frac{\Sigma c_v T}{\sigma T^4}
\end{equation}

\noindent where $c_v$ is the specific heat capacity. Thus
$\tau_\mathrm{cool} \propto \Sigma T^{-3}$ with a very strong
temperature dependance. For an initial value $\Sigma = 350$
g.cm$^{-2}$, $\tau_\mathrm{cool} = 9\ P_\mathrm{orb}$ for $T = 2500$
K but $\tau_\mathrm{cool} = 0.15\ P_\mathrm{orb}$ for $T = 10^4$ K.
Consequently as materials heated at $T \sim 10^4$ K enters the
shadowed region, their very rapid cooling is unavoidable. A more
elaborate computation of a $T^4$ cooling shows that after a few cooling time
scales the gas temperature is close to the ambient value. The 
preceding argument holds even better when one corrects for the Roche
geometry (see Sect. \ref{model}): real distances are about $30\%$
larger.
The cooling of the gas is so efficient that this conclusion does not 
depend much on the thickness of the shadowed region (see discussion at the end 
of Sect. \ref{sectirradiation}) nor on the exact value of
$\Sigma_\mathrm{irr}$, since it would require unrealistically large
values of $\Sigma$ for $\tau_\mathrm{cool}$ to be of the order of
the crossing time scale.

\subsection{Model 2 \& 3: U Gem and Z Cam classes}

Numerical simulations with parameters corresponding to U Gem (model
2) and Z Cam (model 3) show very similar features to the SU Uma
case. In particular the time needed for particles to cross the
shadow boundary is of the same order as previously found, namely a
few orbital periods while the cooling time is unchanged: for the
parameters of model 2 and 3 $\tau_\mathrm{cool} \sim 0.02-0.03\
P_\mathrm{orb}$ (due to larger orbital periods), thus showing that
the above conclusion also applies here.

\subsection{Mass transfer enhancement ?}
\label{sectmasstrenh}

\begin{figure*}[t] %  figure placement: here, top, bottom, or page
\parbox{0.5\linewidth}{\includegraphics[width=0.8\linewidth,angle=90]{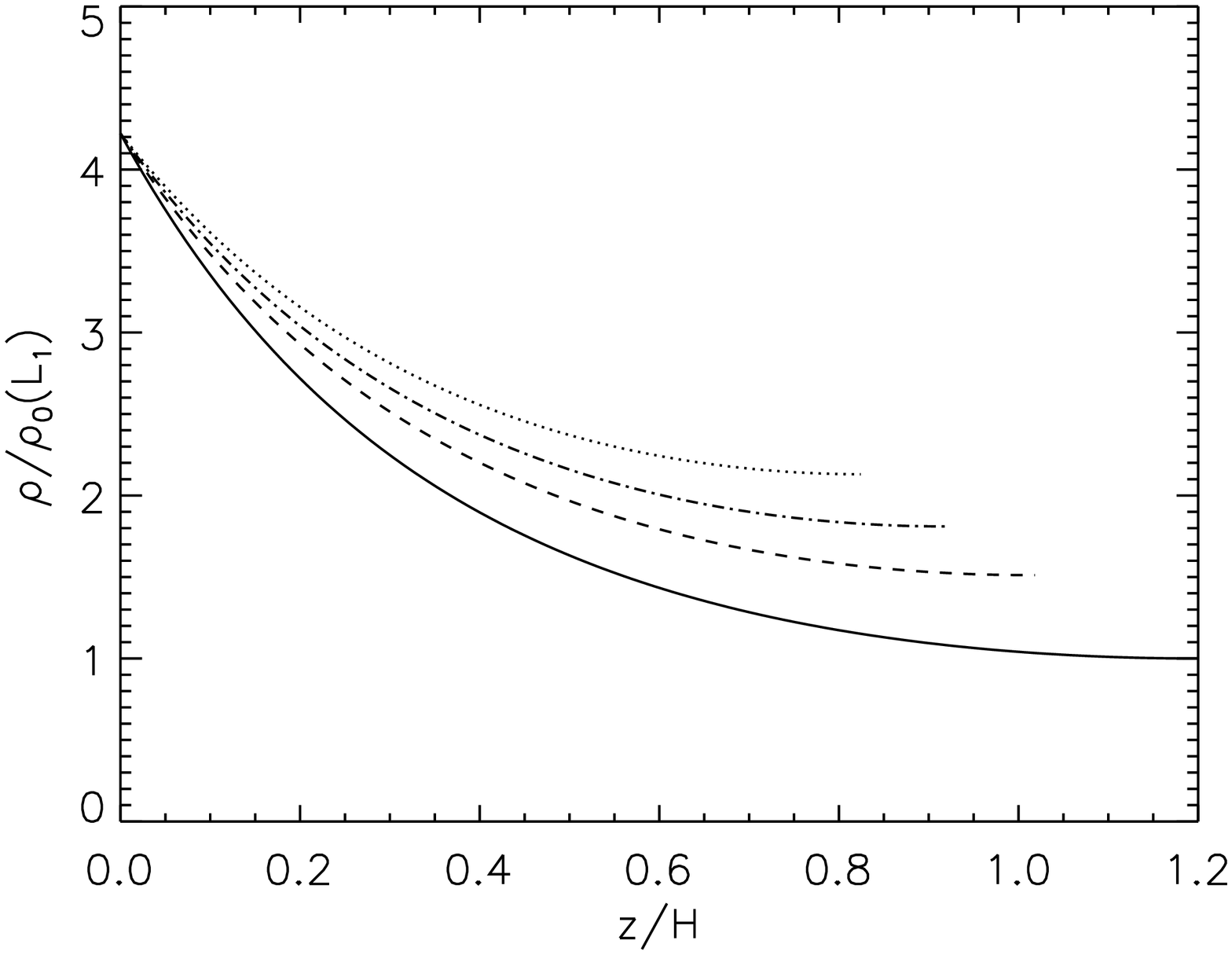}}
\parbox{0.5\linewidth}{\includegraphics[width=0.8\linewidth,angle=90]{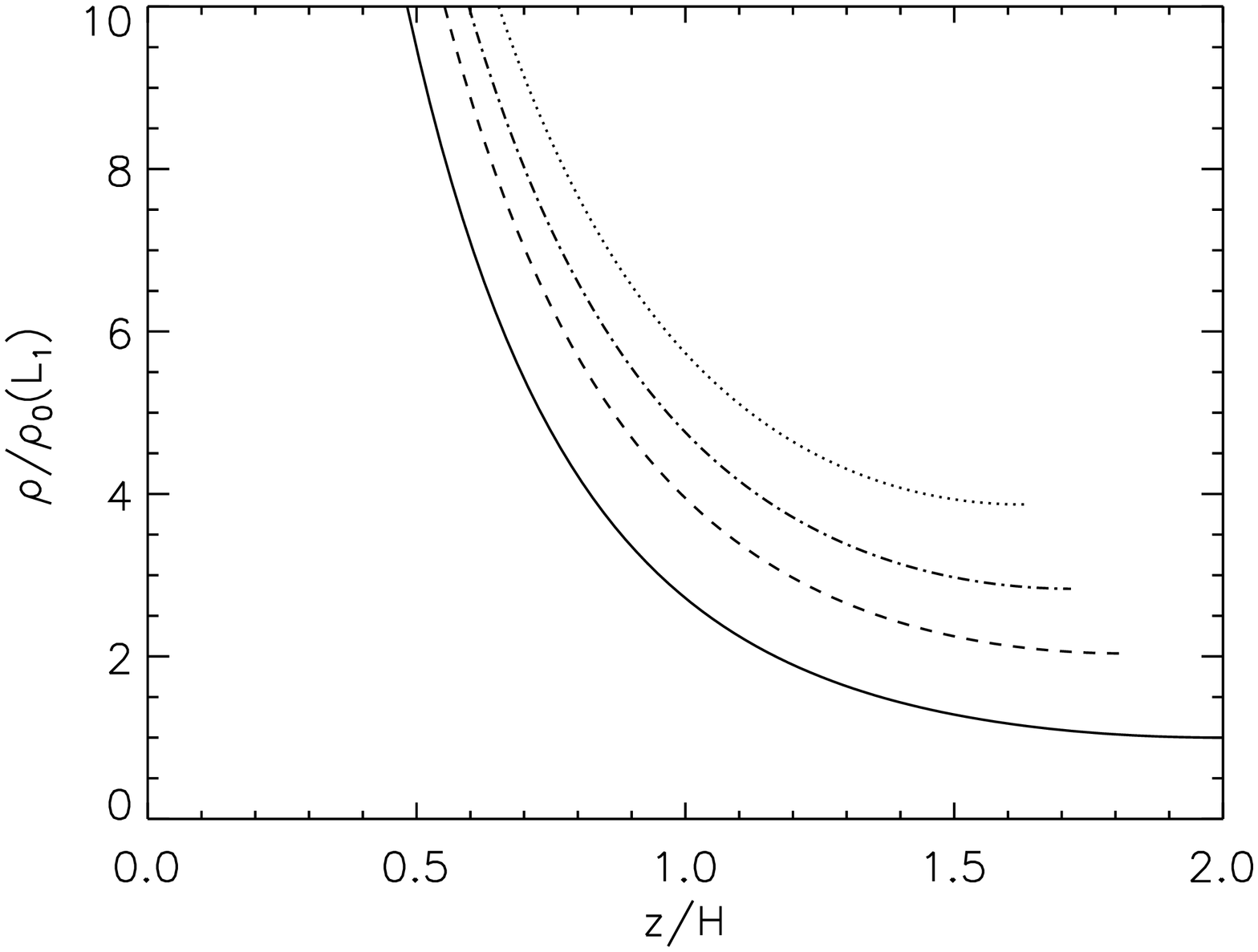}}
   \caption{Vertical profile of the density $\rho$ in the isothermal envelope
   when $v_\theta = 0$, $v_\theta = 0.5 c_\mathrm{s}$, $v_\theta = 0.75
   c_\mathrm{s}$ and $v_\theta = c_\mathrm{s}$ (from bottom to top),
   left panel corresponds to $z_{L_1} = 1.2 H$ and right panel corresponds
   to $z_{L_1} = 2. H$. The density is normalized by the value of the density
   at $L_1$ when $v_\parallel=0$.}
   \label{rho_cor}
\end{figure*}

We now discuss the implications of our numerical results on a
possible mass transfer enhancement. The mass transfer rate is given
by (see Lubow \& Shu 1975, Meyer \& Meyer-Hofmeister 1983):
\begin{equation}
\label{mdot}
\dot M = Q \rho(L_1) c_\mathrm{s}
\end{equation}

\noindent where $c_\mathrm{s}$ is the isothermal sound speed:
\begin{equation}
c_\mathrm{s} = \sqrt{R_gT(L_1)}
\end{equation}

\noindent and $Q$ is the cross section of the stream (see Meyer \&
Meyer-Hofmeister 1983):
\begin{equation}
Q=\frac{2\pi}{k}\big(\frac{c_\mathrm{s}}{\Omega}\big)^2
\label{cross_section}
\end{equation}

\noindent where $k$ is a constant which depends only slightly on the mass ratio $q$.

The factor $Q c_\mathrm{s}$ yields a $T(L_1)^{3/2}$ dependence of
the mass transfer on the temperature at $L_1$. If we further assume
that vertical hydrostatic equilibrium prevails, $\rho(L_1)$ also
depends $T(L_1)$ and on $\Sigma(L_1)$.  Our results have shown that
$T(L_1)$ is almost unaffected by irradiation and heat transport. An
increase of $\Sigma(L_1)$ by a few percents leads to an increase of
$\dot M$ by the same amount and is therefore inefficient to increase
the mass transfer rate (see right panel of Fig. \ref{VL1}). However,
beneath $L_1$ the Coriolis force pushes the gas in the vertical
direction and this could change the vertical distribution of the
mass density. Including the Coriolis force in the hydrostatic
balance gives:

\begin{equation}
\frac{\partial P}{\partial z} = -\rho g + \rho 2\Omega v_\theta = -\rho g_\mathrm{eff}
\label{hydrostatic}
\end{equation}
\noindent where the $z$ axis now refers to the local vertical axis
with the origin taken at the base of the isothermal atmosphere, $g$
is the Roche gravity and $v_\parallel$ is the surface velocity of
the gas.

Due to the Coriolis force, the location where the effective gravity
vanishes can be lower than the $L_1$ point. In this case, one should
replace $\rho(L_1)$ in Eq. \ref{mdot} by the density at the point
where the effective gravity vanishes.

When $v_\parallel = 0$, Eq. \ref{hydrostatic} can be solved by
expanding $g$ to first order around $z_{L_1}$, the height of $L_1$
(see e.g. Lubow and Shu 1975), giving:
\begin{equation}
\rho_0(z) = \rho_0(L_1)\exp((z-z_{L_1})^2/H^2)
\end{equation}
\noindent where $H=c_\mathrm{s}/(\Omega \sqrt{A+1/2})$ is the
vertical scale height of the isothermal envelope ($A$ is a numerical
factor given in Lubow and Shu (1975))

Since we do not known the vertical profile of $v_\parallel$, we
solve Eq. \ref{hydrostatic} with $v_\theta=0.5, 0.75, 1.0\
c_\mathrm{s}$. The results are shown in Fig. \ref{rho_cor}. The mass
transfer enhancement is directly given by the increase of the
density. Results depends on the ratio $z_{L_1}/H$, which has to be
determined by computing the vertical structure of the envelope. This
is beyond the scope of the present paper and we consider in Fig.
\ref{rho_cor} two cases: $z_{L_1}/H=1.2$ and $z_{L_1}/H=2$.

It should be noted that, as the velocity reaches a fraction of the
sound speed, it is possible that the hydrostatic equilibrium
assumption breaks in the whole vertical extend of the envelop. A
detailed analysis would require 3D simulations, out of the scope of
this paper. However, the main conclusion that the cooling time is
much shorter than the crossing time of the disc shadow will not
change.

Finally, note that a mass transfer enhancement due to the strong
surface flow, whatever the mechanism at play, would be restricted to
the rise of the outburst or superoutburst (see left panel of Fig.
\ref{VL1} for $t < 30\ P_\mathrm{orb}$).

\section{Conclusion}
\label{conclusion}

We have numerically investigated the surface flow of the irradiated
secondary star in dwarf novae, using a simple model for both the
geometry and irradiation of the secondary star.

Our numerical simulations can be viewed as large scale
meteorological simulations of irradiated secondary stars:
irradiation triggers a large anti-cyclonic perturbation (i.e. high
pressure perturbation) with the associated clockwise circulation due
to the Coriolis force. An important result of our numerical
simulations is that the Roche geometry naturally leads to a
circulation flow transiting by $L_1$, contrary to naive
expectations. This provides a way to transport matter/heat in the
vicinity of $L_1$, where matter leaves the secondary. We have also
shown that it is necessary to solve the full time-dependent problem
in order to catch the dynamics of the geostrophic adjustment
process; the flow does not reach the geostrophic state without
modifying the background pressure gradients (as is implicitly done
in Osaki \& Meyer 2003; 2004). It is also important to note that no
steady state can be reached on the time scale of an outburst or a
superoutburst.

We modeled the irradiation and radiative cooling of the envelope of
the secondary in a very crude way, thus avoiding the necessity of
solving the vertical structure of the atmosphere. In our model,
radiative cooling occurs in a very short time scale, much shorter
than the crossing time scale of the shadowed region. There is thus
no efficient heat transport in the vicinity of $L_1$. It is unlikely
that this conclusion will change in a 3D approach that would also
include a more realistic treatment of radiative transfer. As a
consequence, if hydrostatic equilibrium prevails, the only effect
that could lead to a moderate enhancement of mass transfer is the
vertical Coriolis force. It could however be possible that
hydrostatic equilibrium breaks down as the magnitude of the surface
flow becomes of the order of the sound speed. A 3D approach would
then be necessary to determine the mass flux leaving the secondary.
Note that we have not considered here the magnetic field of the
secondary. For the expected range of intensity, the magnetic field
is unlikely to dominate the gas dynamics but its influence on the
flow could be of some importance. An correct implementation of the
magnetic field configuration would however be very difficult and is
far beyond the scope of the paper.

Apart from the transport mechanism investigated here, any other
mechanisms that could directly heat the $L_1$ point would also
contribute to an increase of the mass transfer rate. For example the
direct heating of the $L_1$ point by the rim of the accretion disc
or by the scattering of the accretion luminosity by outflowing
matter are currently under investigation to check if heating of
$L_1$ could be efficient. We leave this for a forthcoming paper.

Even if the circulation flow found here were to lead to a
significant increase of the mass transfer rate, this effect would
not last longer that $10-20$ orbital periods, as the anti-cyclonic
perturbation moves rapidly westwards, quenching the flow through
$L_1$ on a time scale comparable to the rise time of the outburst.
We would expect in this case no differences between outbursts and
superoutbursts.

Finally, it could be tempting to apply these results to soft X-ray
transients (see e.g. Chen et al. 1997 for a review). In these
systems the irradiation flux is larger than in DN by a factor of
order $10^3$ and consequently $T_\mathrm{irr} \sim 5.10^4$ K. Such a
strong irradiation raises very important pressure gradients and one
could expect supersonic velocities. A numerical investigation of
this case would need a more robust numerical code.

\end{document}